\documentclass[a4paper,11pt]{article}
\pdfoutput=1 

\usepackage{jheppub} 

\usepackage[T1]{fontenc} 

\usepackage{amsmath,amssymb,amsthm,graphicx}
\usepackage[dvipsnames]{xcolor}
\usepackage{multirow}
\usepackage{adjustbox}
\usepackage{comment}
\usepackage{enumerate}
\usepackage{hyperref}
\usepackage{cleveref}
\usepackage{bm}
\usepackage{subcaption}
\usepackage{lscape}
\usepackage{multirow}
\usepackage{booktabs}
\usepackage{adjustbox}

\title{
A Unified Geometric Framework for Understanding Collider Event Manifolds
}

\author[a,b]{Tianji Cai,}
\author[c]{Hancheng Li}

\affiliation[a]{School of Physical Science and Engineering, Tongji University, Shanghai 200092, China}
\affiliation[b]{State Key Laboratory of Autonomous Intelligent Unmanned Systems, MOE Frontiers Science Center for Intelligent Autonomous Systems, Tongji University, Shanghai 200092, China}

\affiliation[c]{Department of Physics and Astronomy, Rutgers, the State University of New Jersey, 126 Frelinghuysen Road, Piscataway, NJ 08854 USA}

\emailAdd{tianji\_cai@tongji.edu.cn}
\emailAdd{hl1417@physics.rutgers.edu}

\abstract{
By defining notions of similarity between collider events, metrics give rise to event manifolds with nontrivial geometric structure. Understanding these structures and their relation to the underlying physics remains largely unexplored, partly due to the challenge of systematically comparing spaces endowed with distinct metrics. We introduce the multi-reference relative representation (M3R) framework, which provides a unified coordinate system for analyzing different spaces on equal footing. Applying M3R to three physically motivated event metrics---the phase space, spectral 2-Wasserstein, and energy-flow 2-Wasserstein metrics---we study individual event manifolds and the decision boundaries separating different physical processes. We find that these metrics resolve distinct yet partially shared structures, while intra- and inter-manifold geometries exhibit qualitatively different properties. Combining multiple metrics further reveals complementary information that can be exploited within a unified representation. More broadly, M3R provides a general language for comparing and composing different notions of event similarity, opening a path toward a systematic geometric description of collider event space.
}

\begin{document} 
\maketitle
\flushbottom

\newpage

\section{Introduction} \label{sec:intro}

Collider events populate a highly complex space whose rich structure reflects the underlying dynamics of particle interactions. Quantitatively characterizing this space requires a notion of distance that measures the similarity between events. A physically meaningful metric endows collider event space with an intrinsic geometry~\cite{PhysRevLett.123.041801}, allowing the resulting \textit{event manifold} to be studied as an object in its own right and revealing structure beyond that captured by any prescribed set of observables~\cite{Komiske_2020}. Such metric constructions have enabled a broad range of applications in collider physics, including jet and event classification, anomaly detection~\cite{Fraser:2021lxm, Craig:2024rlv, Cai:2025vxl, dagnolo2025intrinsicdimensioncolliderevents}, event reweighting~\cite{Andersen:2026ppe, Doherty:2026fli, Hay:2026tbh}, construction of new observables~\cite{Cesarotti:2020hwb, Ba:2023hix, ATLAS:2023mny, Cesarotti:2020xtf, Gambhir:2024ndc}, and quantitative comparisons between theories~\cite{Komiske_2020, Jain:2026meq}.

There is, however, no unique geometry of collider event space. A variety of physically motivated metrics have been developed, including optimal transport and phase space constructions~\cite{PhysRevD.102.116019, PhysRevD.105.076003, cai2025multiscaleoptimaltransportcomplete, Larkoski:2023qnv, PhysRevD.102.094014, Cai2025phasespace}. Built from different physical and mathematical principles, these metrics encode distinct notions of event similarity and can therefore induce different geometric structures on the same ensemble of events. While individual metrics have been extensively studied~\cite{Larkoski:2025clo, Larkoski:2025idq, 6r43-xhtj}, there remains a conceptual gap between the definition of particular metrics and a systematic understanding of the geometries they induce. A common framework is therefore needed to place heterogeneous collider metrics on the same footing and systematically characterize the resulting geometric spaces.

Addressing this gap presents both conceptual and practical challenges. Relating the distinct geometries induced by different metrics while retaining direct access to their underlying physical structure is nontrivial. General mathematical tools for comparing metric spaces are well established, including methods based on global geometric discrepancies, cross-space correspondences, and topological invariants~\cite{Gromov:2007, Memoli:2011, Chazal:2009}. Instead, our goal is to retain an event-level description of each geometry, allowing us to probe which structures are resolved by a given metric, how these structures depend on the underlying physical process, and how they contribute to distinguishing different processes. This conceptual challenge is compounded by a practical one: many collider metrics are computationally expensive, and resolving the geometry of an ensemble of $N$ events through all pairwise distances requires $\mathcal{O}(N^2)$ metric evaluations. The cost quickly becomes prohibitive for large event samples and multiple metrics. A useful framework should therefore provide both a physically interpretable common description of event geometries and a scalable alternative to exhaustive pairwise metric evaluation.

To this end, we introduce the Multi-Reference Relative Representation (M3R) framework, which characterizes each event through its distances to a finite set of reference events, providing a unified finite-dimensional representation for different collider geometries. Unlike global comparisons between metric spaces, related landmark-based and dissimilarity methods have been developed to obtain object-level relative representations for pattern recognition, manifold characterization, and, more recently, alignment across learned representation spaces~\cite{PekalskaDuin:2005, LiangPaisley:2015, moschella2023relativerepresentationsenablezeroshot}. M3R extends this perspective to physically defined collider geometries, using relative representations not only as an embedding strategy but as a tool for probing how different physical metrics organize collider data.

The M3R framework enables us to explore geometric questions that are difficult to formulate otherwise. In particular, we distinguish two different objects: an individual event manifold, which characterizes how events of the same underlying physical process are organized, and the decision boundary between event manifolds, which reflects how different physical processes are separated under a given notion of event similarity. Together, these provide two key perspectives on how collider metrics encode the underlying physics, through both the internal structure of event manifolds and their separation from one another.

Throughout this work, we adopt an operational approach: rather than pursuing an abstract or exhaustive description of the underlying metric spaces, we probe their geometry through physically interpretable and computationally accessible quantities that can be constructed from finite collider datasets. This perspective allows notions such as geometric complexity and structural similarity to be operationally characterized through concrete analysis procedures, while keeping them closely connected to the physical information encoded by the metrics and their practical role in collider analysis.

The remainder of this paper is organized as follows. Section~\ref{sec:metrics} reviews the three collider event metrics considered in this work, together with event dissimilarities incorporating non-kinematic information. Section~\ref{sec:m3r} introduces the M3R framework and the analysis setup. Sections~\ref{sec:single-manifold} and~\ref{sec:decision-boundary} study, respectively, the geometry of individual event manifolds and the inter-manifold decision boundaries. Section~\ref{sec:unified-manifolds} then combines complementary notions of event similarity into a unified representation. We conclude in section~\ref{sec:conclusions} with a discussion of broader implications and future directions.

\section{Collider Event Metrics} \label{sec:metrics}

A metric endows the space of collider events with a geometry by assigning a distance to each pair of events. Different choices of metric probe different aspects of event structure and can therefore give rise to distinct geometries on the same event space. In this section, we briefly review the three event metrics: the energy-flow 2-Wasserstein metric, the spectral 2-Wasserstein metric, and the relativistic $N$-body phase space metric. We focus on their definitions, physical interpretations, and implementations used in the subsequent analysis; detailed treatments can be found in the original works~\cite{PhysRevD.102.116019, Larkoski:2023qnv, Gambhir:2024ndc, PhysRevD.102.094014, Cai2025phasespace}.

\subsection{2-Wasserstein Metric on Event Energy Flow}
\label{subsec:w2-ef}

Optimal transport (OT) provides a natural construction of distances between distributions by minimizing the cost of transporting one distribution into another~\cite{peyré2020computationaloptimaltransport}. The resulting distance depends on the ground space on which the distributions are defined and the ground metric that specifies the transport cost. In this work, we consider two OT-based event metrics, the 2-Wasserstein metric (section~\ref{subsec:w2-ef}) and the spectral 2-Wasserstein metric (section~\ref{subsec:w2-sp}), which share the same optimal transport construction but differ in the representation of an event and the associated ground space.

For the 2-Wasserstein (W$_2$) metric, a collider event\footnote{Throughout this work, an ``event'' denotes the object on which a metric is defined and may refer to either a full collision event or an individual jet. The numerical studies are all performed on jets.} is represented by its energy flow on the rapidity--azimuth plane, which defines the ground space. An event $\mathcal{E}$ with $M$ constituent particles is therefore represented by the discrete measure
\begin{equation}
    \mathcal{E} = \sum_{i=1}^{M} z_i\,\delta_{x_i},
    \qquad
    z_i \equiv \frac{p_{T,i}}{\sum_k p_{T,k}},
    \qquad
    \sum_i z_i = 1,
    \label{eq:energy_flow}
\end{equation}
where $x_i=(y_i,\phi_i)$ specifies the location of the $i$th constituent in the ground space $\Omega \subset \mathbb{R}\times S^1$. The ground metric is the Euclidean distance between two constituents $i$ and $j$, i.e.,
\begin{equation}
    d_{\rm ground}(x_i,x_j)
    \equiv \Delta R_{ij}
    =
    \sqrt{(y_i-y_j)^2+(\phi_i-\phi_j)^2},
    \label{eq:w2_ground_metric}
\end{equation}
with periodicity in $\phi$ understood.

The 2-Wasserstein distance between two events $\mathcal{E}=\sum_i z_i\delta_{x_i}$ and $\mathcal{E'}=\sum_j z_j'\delta_{x_j'}$ is then defined as
\begin{equation}            
    \text{W}_2(\mathcal{E},\mathcal{E'})
    =
    \left[
    \min_{\gamma\in\Gamma(\mathcal{E},\mathcal{E'})}
    \sum_{i,j}\gamma_{ij}\,
    d_{\rm ground}(x_i,x_j')^2
    \right]^{1/2},
    \label{eq:w2}
\end{equation}
where
\begin{equation}
    \Gamma(\mathcal{E},\mathcal{E'})
    =
    \left\{
    \gamma_{ij}\geq 0
    \,\middle|\,
    \sum_j\gamma_{ij}=z_i,\quad
    \sum_i\gamma_{ij}=z_j'
    \right\}
    \label{eq:w2_transport_plan}
\end{equation}
is the set of admissible transport plans. Each $\gamma_{ij}$ specifies the $p_T$ weight transported from constituent $i$ in event $\mathcal{E}$ to constituent $j$ in event $\mathcal{E'}$. The W$_2$ metric thus lifts the Euclidean distance on the ground space $\Omega$ to a distance between event energy flows by minimizing the total transport cost over all admissible rearrangements.

Physically, W$_2$ accounts for both the amount of transverse momentum being redistributed and the distance in the $y-\phi$ plane over which it is transported. Local rearrangements of $p_T$ incur a small cost, whereas transporting a comparable amount of $p_T$ over large $y-\phi$ separations is penalized more strongly. The resulting geometry therefore directly resolves the spatial organization of the event energy flow, while the normalization to unit total transverse momentum makes it independent of the overall event $p_T$ scale. 

W$_2$ is invariant under constituent permutations and is infrared and collinear (IRC) safe,\footnote{The W$_2$ metric is IRC safe when applied to normalized energy flows~\cite{PhysRevLett.123.041801, PhysRevD.102.116019}. Under the addition of a particle carrying a fraction $\epsilon$ of the event transverse momentum, one can explicitly construct a transport plan satisfying $\mathrm{W}_2(\mathcal E_\epsilon,\mathcal E)=\mathcal O(\sqrt{\epsilon})$. Similarly, for a collinear splitting $k\to k_1+k_2$, transporting the daughter weights back to the parent position gives $\mathrm{W}_2(\mathcal E_{\rm split},\mathcal E)\to0$ in the collinear limit.} making it well suited for collider events with variable constituent multiplicities and soft/collinear radiation. Among the family of Wasserstein distances, the $p=2$ case considered here is further distinguished by the formal Riemannian structure of the Wasserstein space, which enables a tangent space description and motivates the linearized W$_2$ representation of collider events~\cite{PhysRevD.102.116019}.

\subsection{2-Wasserstein Metric on Event Spectral Function}
\label{subsec:w2-sp}

The spectral Wasserstein construction provides a complementary optimal transport geometry for the space of collider events. Whereas the energy-flow W$_2$ metric compares events directly as distributions of transverse momentum on the rapidity--azimuth plane, the spectral construction instead represents the same event through its spectral function, a one-dimensional distribution encoding pairwise angular separations weighted by products of constituent momenta. The 2-Wasserstein distance is then evaluated between these spectral functions, replacing transport on the two-dimensional $y-\phi$ plane with transport on a one-dimensional spectral ground space. The resulting transport problem admits an exact closed-form solution and has enabled the construction of a broad class of IRC-safe jet and event shape observables.

More precisely, for an event $\mathcal{E}$ with $M$ constituents, each with transverse-momentum fraction $z_i$, we define its spectral function as
\begin{align}
    s_{\mathcal{E}}(\omega)
    &=
    \sum_{i,j=1}^M z_i z_j\,
    \delta\!\left(\omega-\omega_{ij}\right)
    \nonumber\\
    &=
    \sum_{i=1}^M z_i^2\,\delta(\omega)
    +
    2\sum_{i<j}^M z_i z_j\,
    \delta\!\left(\omega-\omega_{ij}\right).
    \label{eq:spectral-measure}
\end{align}
Here $i$ and $j$ label constituents of the same event $\mathcal{E}$, and $\omega_{ij} \equiv \Delta R_{ij}$ is their pairwise distance on the $y-\phi$ plane, as defined in eq.~\ref{eq:w2_ground_metric}. The spectral function therefore represents an event as a one-dimensional distribution over its internal angular scales: each pair of constituents contributes spectral weight $2z_i z_j$ at their separation $\omega_{ij}$, while the self-terms contribute $\sum_i z_i^2$ at $\omega=0$. With the normalization $\sum_i z_i=1$, the spectral function has unit total weight,
\begin{equation}
    \int_0^\infty d\omega\,s_{\mathcal{E}}(\omega)
    =
    \left(\sum_i^M z_i\right)^2
    =1. \label{eq:spectral_normalization}
\end{equation}
Further, it is manifestly invariant under constituent permutations and isometries of the underlying angular ground space, as well as being IRC safe.\footnote{In the soft limit, all additional spectral weights involving the soft constituent vanish with its $p_T$ fraction. Under an exactly collinear splitting $i\to a+b$ with $z_i=z_a+z_b$, the self- and daughter-daughter contributions at $\omega=0$ combine to $z_a^2+z_b^2+2z_a z_b=(z_a+z_b)^2=z_i^2$, while the correlations with every other constituent $j$ combine as $2(z_a+z_b)z_j=2z_i z_j$. The spectral function is therefore unchanged in the exact collinear limit.}

For two events $\mathcal{E}$ and $\mathcal{E}'$, define the corresponding cumulative spectral functions as 
\begin{equation}
S_\mathcal{E}(\omega)
=
\int_0^\omega d\tilde{\omega}
s_{\mathcal{E}}(\tilde{\omega}),
\qquad
S_{\mathcal{E}'}(\omega)
=
\int_0^\omega d\tilde{\omega}
s_{\mathcal{E}’}(\tilde{\omega}).
\label{eq:cumulative_spectral_function}
\end{equation}
Because these are discrete measures, we can use their generalized inverse, or quantile functions,
\begin{equation}
    S_\mathcal{E}^{-1}(u)
    =
    \inf\left\{
    \omega\geq0
    \,\middle|\,
    S_\mathcal{E}(\omega)\geq u
    \right\},
    \qquad
    u\in[0,1],
\label{eq:inverse_spectral_function}
\end{equation}
and analogously for $S_{\mathcal{E}'}^{-1}$, to compute the spectral 2-Wasserstein (SW$_2$) distance
\begin{equation}
\mathrm{SW}_2(\mathcal{E},\mathcal{E}’)
=
\left[
\int_0^1 du 
\left|
S_\mathcal{E}^{-1}(u)-S_{\mathcal{E}'}^{-1}(u)
\right|^2
\right]^{1/2}.
\label{eq:sw2}
\end{equation}
Unlike higher-dimensional optimal transport, one-dimensional OT problems admit an explicit quantile-matching solution. For discrete collider events, the SW$_2$ distance can be evaluated from the ordered pairwise angular separations and their associated weights, without solving a multidimensional transport problem.

\begin{figure}[htbp]
    \centering
    \includegraphics[width=0.49\linewidth]{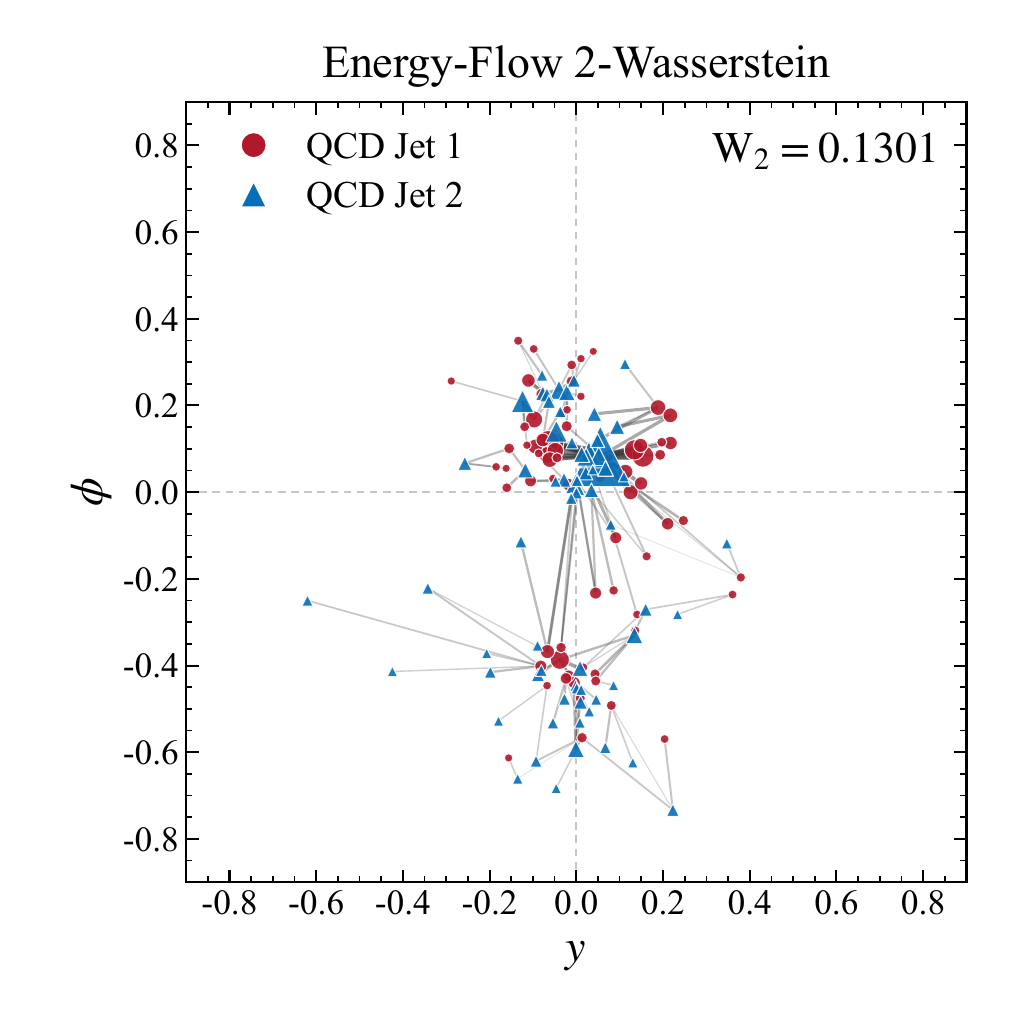}
    \hfill
    \includegraphics[width=0.49\linewidth]{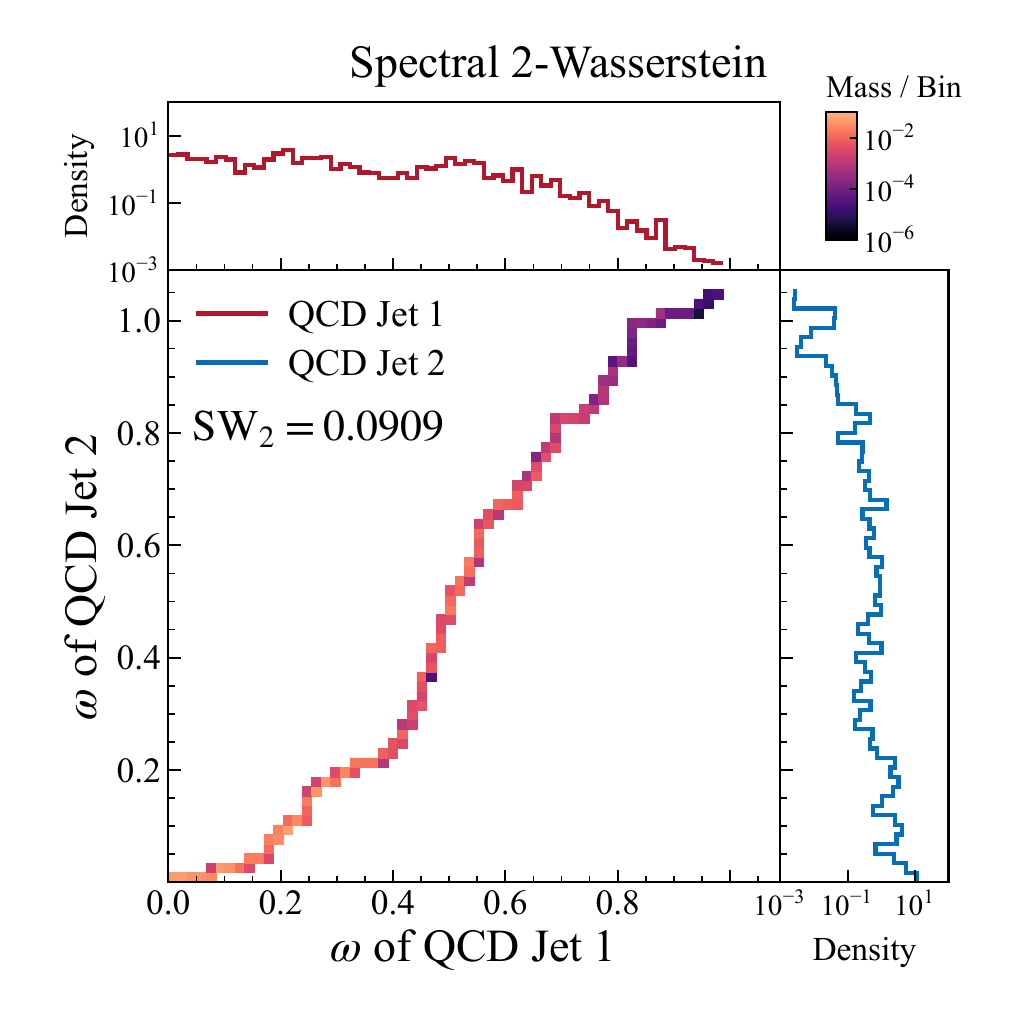}
    \caption{
        Optimal transport plans between the same pair of QCD jets under the 2-Wasserstein (W$_2$, left) and spectral 2-Wasserstein (SW$_2$, right) distances. \textbf{Left:} Jet constituents in the rapidity--azimuth plane, with marker area proportional to the constituent $p_T$ fraction. Lines indicate nonzero optimal couplings, with width and opacity proportional to the transported weight. \textbf{Right:} Optimal coupling between the corresponding spectral functions, with their marginal densities shown along the top and right axes. The resulting distances are quoted in each panel.
    }
    \label{fig:w2_sw2_plan}
\end{figure}

Figure~\ref{fig:w2_sw2_plan} illustrates the optimal transport plans under W$_2$ and SW$_2$ for the same pair of QCD jets, highlighting their distinct ground spaces and transport structures. Nevertheless, the two distances are closely related for simple few-particle configurations. In the soft and collinear limits, W$_2$ and SW$_2$ coincide up to an overall normalization for the one-emission configurations appearing at $\mathcal{O}(\alpha_s)$, with the correspondence extending to certain higher-multiplicity configurations at $\mathcal{O}(\alpha_s^2)$~\cite{Larkoski:2023qnv}. More generally, when one of the two events consists of a single constituent particle, the W$_2$ and SW$_2$ distances are proportional~\cite{Gambhir:2024ndc}.
This suggests that W$_2$ and SW$_2$ should resolve similar geometric structure in event samples dominated by strongly ordered soft and collinear radiation.

This correspondence, however, does not extend to general multiparticle configurations. By retaining only the weighted distribution of pairwise angular separations, the spectral representation can map distinct events to the same spectral function, with such degeneracies arising already for three-particle configurations. SW$_2$ and W$_2$ therefore resolve different event geometries for nontrivial multi-prong structures, as observed numerically for top jets~\cite{Gambhir:2024ndc}.

These results motivate our systematic comparison of the two metric geometries across different jet classes. In particular, we expect stronger correspondence for QCD jets dominated by soft and collinear radiation, while more complex multi-pronged jets can expose differences between the information retained by the energy-flow and spectral representations. We test these expectations directly in section~\ref{sec:single-manifold}.

\subsection{Metric on Relativistic $N$-body Phase Space}
\label{subsec:ps}

Beyond optimal transport based constructions, a notion of distance between collider events can be derived directly from the intrinsic geometry of relativistic $N$-body phase  space, on which scattering amplitudes are defined. For $N$ massless final-state particles at fixed center-of-mass energy, the phase space manifold is isomorphic to the product space of an $(N-1)$-dimensional simplex and a $(2N-3)$-dimensional hypersphere, $\Pi_N \simeq \Delta^{N-1}\times S^{2N-3}$~\cite{PhysRevD.102.094014}. Building on this covariant description, $\Pi_N$ can be equipped with explicit global coordinates and a practical Riemannian metric~\cite{Cai2025phasespace}. The resulting phase space (PS) distance for collider events has demonstrated its utility in both full-event and single-jet classification tasks and has subsequently been adopted in experimental CMS searches for microscopic black holes, string balls, and electroweak sphalerons~\cite{ CMS2026}.

The phase space construction applies generally to multiparticle collider events, with its precise realization depending on the kinematic regime under consideration. The numerical studies in this work, introduced in section~\ref{subsec:dataset}, concern individual boosted jets rather than complete collision events. We therefore employ the collinear formulation of the PS distance, appropriate for a collection of highly energetic particles localized around an axis. Below we summarize only the ingredients relevant to this implementation and refer to Refs.~\cite{PhysRevD.102.094014, Cai2025phasespace} for the general construction and detailed derivation.

For a system with $N$ massless particles of four-momenta $\{p_k^\mu\}$, we first choose the total momentum of the system to define the $+z$ direction. The simplex coordinates are then given by the normalized light-cone momentum components
\begin{equation}
    \rho_k =
    \frac{p_k^0-p_k^3}
    {\sum_{j=1}^{N}(p_j^0-p_j^3)},
    \qquad
    \sum_{k=1}^{N}\rho_k=1.
    \label{eq:ps_rho}
\end{equation}
In other words, the coordinates $\boldsymbol{\rho}=(\rho_1,\ldots,\rho_N)$ lie on the simplex $\Delta^{N-1}$. The remaining degrees of freedom are encoded in an $(N-1)$-component complex unit vector $\tilde{\boldsymbol{v}}\in S^{2N-3}$, obtained from the particle four-momenta through a coordinate transformation. Together, $(\boldsymbol{\rho},\tilde{\boldsymbol{v}})$ provide global coordinates for the $N$-body phase space.

For a boosted system, the collinear construction additionally introduces the dimensionless variable
\begin{equation}
    \chi = \frac{p^0-p^3}{p^0+p^3},
    \qquad
    \text{with}
    \quad
    p^0=\sum_{k=1}^N p_k^0,
    \quad
    p^3=\sum_{k=1}^N p_k^3,
    \label{eq:ps_chi}
\end{equation}
that is, $p^0$ and $p^3$ are the total energy and longitudinal momentum of the system, respectively. In the collinear limit, $\chi$ characterizes the overall opening
scale of the system with respect to its boost direction, with $\log\chi$ providing the natural coordinate associated with this degree of freedom.

The distance between two boosted jets $\mathcal{J}$ and $\mathcal{J}'$ can consequently be decomposed into contributions from the simplex, the hypersphere, and the collinear scale factor. We therefore have
\begin{equation}
    d_\Delta(\mathcal{J},\mathcal{J}')
    =
    \left[
        \sum_{k=1}^{N}
        \left(\rho_{k,\mathcal{J}}-\rho_{k,\mathcal{J}'}\right)^2
    \right]^{1/2},
   \label{eq:ps_simplex_distance}
\end{equation}
for the simplex component, and
\begin{equation}
    d_\chi(\mathcal{J},\mathcal{J}')
    =
    \left|\log\chi_{\mathcal{J}}-\log\chi_{\mathcal{J}'}\right|
    \label{eq:ps_chi_distance}
\end{equation}
for the collinear scale factor. The hypersphere contribution is given by the geodesic distance on $S^{2N-3}$. Accounting for the physically irrelevant overall azimuthal orientation and reflection of the event, this can be written as
\begin{equation}
    d_S(\mathcal{J},\mathcal{J}')
    =
    \cos^{-1}\!\left[
        \max\left(            \left|\tilde{\boldsymbol{v}}^{\,\dagger}_{\mathcal{J}}      \tilde{\boldsymbol{v}}_{\mathcal{J}'}\right|,            \left|\tilde{\boldsymbol{v}}^{\,T}_{\mathcal{J}}  \tilde{\boldsymbol{v}}_{\mathcal{J}'}\right|
        \right)
    \right].
    \label{eq:ps_sphere_distance}
\end{equation}
The full PS distance is then constructed as a weighted combination of these geometrically distinct contributions,
\begin{equation}
    d_{\rm PS}(\mathcal{J},\mathcal{J}')
    =
    \left[
        d_\Delta^2(\mathcal{J},\mathcal{J}')
        +\lambda_S d_S^2(\mathcal{J},\mathcal{J}')
        +\lambda_\chi d_\chi^2(\mathcal{J},\mathcal{J}')
    \right]^{1/2},
    \label{eq:ps_distance}
\end{equation}
up to an overall normalization factor determined by the fixed phase space volume element. We use $\lambda_S=1$ and $\lambda_\chi=16\pi^2$ throughout the analysis.

A final practical consideration concerns the fixed multiplicity and constituent permutation invariance required by the phase space distance definition. Since reconstructed jets generally contain a variable number of constituents, we select only jets with at least 20 constituent particles and recluster them into $N=10$ subjets using the exclusive $k_T$ algorithm with $R=1.0$. The phase space formalism further assumes massless constituents; we impose this condition by manually setting $p_k^0=|\vec{p}_k|$ for each subjet. The resulting subjets are then ordered by decreasing $p_T$, which establishes a common correspondence between the subjets of different jets and, in particular, guarantees that identical jets are represented identically. We use this prescription as an approximation to the exact minimization over $N!$ subjet permutations.

The PS metric provides a qualitatively different notion of event similarity from the two transport-based metrics. Whereas W$_2$ and SW$_2$ represent an event as a distribution on an underlying ground space and compare two events through the optimal transport between these distributions, the phase space construction represents each event as a point on a fixed-dimensional relativistic phase space manifold and measures distances directly through its intrinsic geometry. These distinct constructions probe different organizations of multiparticle kinematics, motivating the systematic comparison of the event geometries they induce.

\subsection{Incorporating Non-kinematic Information}
\label{subsec:non-kin}

The three event metrics are constructed entirely from particle kinematics. Collider events, however, also contain important non-kinematic information, such as particle identity and electric charge. A prominent example is the observable jet charge~\cite{Krohn_2013}, constructed from the electric charges of jet constituents and typically weighted by their momenta, which is shown to be sensitive to the charge and flavor of the initiating parton and has applications in both Standard Model measurements and searches for new particles~\cite{PhysRevD.93.052003, CMS2017}. Similarly, charged-particle multiplicity provides powerful discrimination between quark- and gluon-initiated jets and probes both perturbative radiation and nonperturbative fragmentation~\cite{PhysRevLett.107.172001, ATLAS2016}. The M3R framework offers a natural way to incorporate such information alongside kinematic metrics without modifying the metrics themselves. Additional event-level quantities can be equipped with their own notions of similarity and subsequently combined within the common relative representation introduced in section~\ref{sec:m3r}.

As a simple demonstration, we make use of the per-particle identification (PID) information commonly provided in benchmark datasets. Each constituent is assigned to one of five particle categories: electron, muon, photon, charged hadron, or neutral hadron. In addition, the electric charge of each constituent is given. For each particle category $a$, we define the corresponding event-level multiplicity as
\begin{equation}
    N_a(\mathcal{E})
    =
    \sum_{i\in\mathcal{E}}
    \mathbf{1}_{a}(i),
    \qquad
    a\in \{e,\mu,\gamma,h^{\pm}, h^0\},
    \label{eq:pid_multiplicity}
\end{equation}
where $\mathbf{1}_{a}(i)$ is one if constituent $i$ belongs to category $a$ and zero otherwise. Similarly, the total electric charge of the event is
\begin{equation}
    Q(\mathcal{E})
    =
    \sum_{i\in\mathcal{E}} q_i,
    \label{eq:jet_charge}
\end{equation}
where $q_i$ denotes the electric charge of each constituent. These six scalars characterize the non-kinematic particle content of each event, i.e.,
\begin{equation}
    \mathbf{h}_{\rm PID}(\mathcal{E})
    =
    \left(
    N_e,N_\mu,N_\gamma,N_{h^{\pm}},N_{h^0},Q
    \right).  \label{eq:nonkinematic_features}
\end{equation}
For each component $h_a$, we define a simple distance between two events $\mathcal{E}$ and $\mathcal{E}'$ by the absolute difference
\begin{equation}
    d_a(\mathcal{E},\mathcal{E}')
    =
    \left|
    h_a(\mathcal{E})-h_a(\mathcal{E}')
    \right|.    \label{eq:nonkinematic_distance}
\end{equation}

Each $d_a$ supplies an independent notion of event similarity associated with a particular aspect of particle content. Unlike the kinematic metrics, these distances do not encode the spatial distribution or momentum configuration of the constituents, and is in general not IRC safe. In particular, the multiplicity distances are manifestly infrared unsafe, since the addition of an arbitrarily soft particle changes the corresponding $N_a$ by one, and are likewise collinear unsafe. These distances should therefore be viewed as phenomenological measures of particle content, complementary to the IRC-safe kinematic metrics. In section~\ref{sec:unified-manifolds}, we construct a unified representation from the three kinematic metrics and the six non-kinematic distances, providing a controlled setting in which to study whether particle species supply complementary discriminatory information beyond that encoded by event kinematics alone.

\section{The Multi-Reference Relative Representation (M3R) Framework}
\label{sec:m3r}

Having reviewed the different collider event metrics, we now establish a common framework for their systematic analysis. We introduce the Multi-Reference Relative Representation (M3R) framework, which characterizes each event through its distances to a finite set of reference events, thereby mapping different metric geometries into a common representation space. In section~\ref{subsec:framework}, we formulate the M3R construction and discuss its geometric and physical properties. Section~\ref{subsec:dataset} then introduces the \texttt{JetClass} dataset used as the common benchmark for the numerical analysis throughout this work.

\subsection{Framework Construction and Properties}
\label{subsec:framework}

\begin{figure}[htbp]
    \centering
    \includegraphics[width=\linewidth]{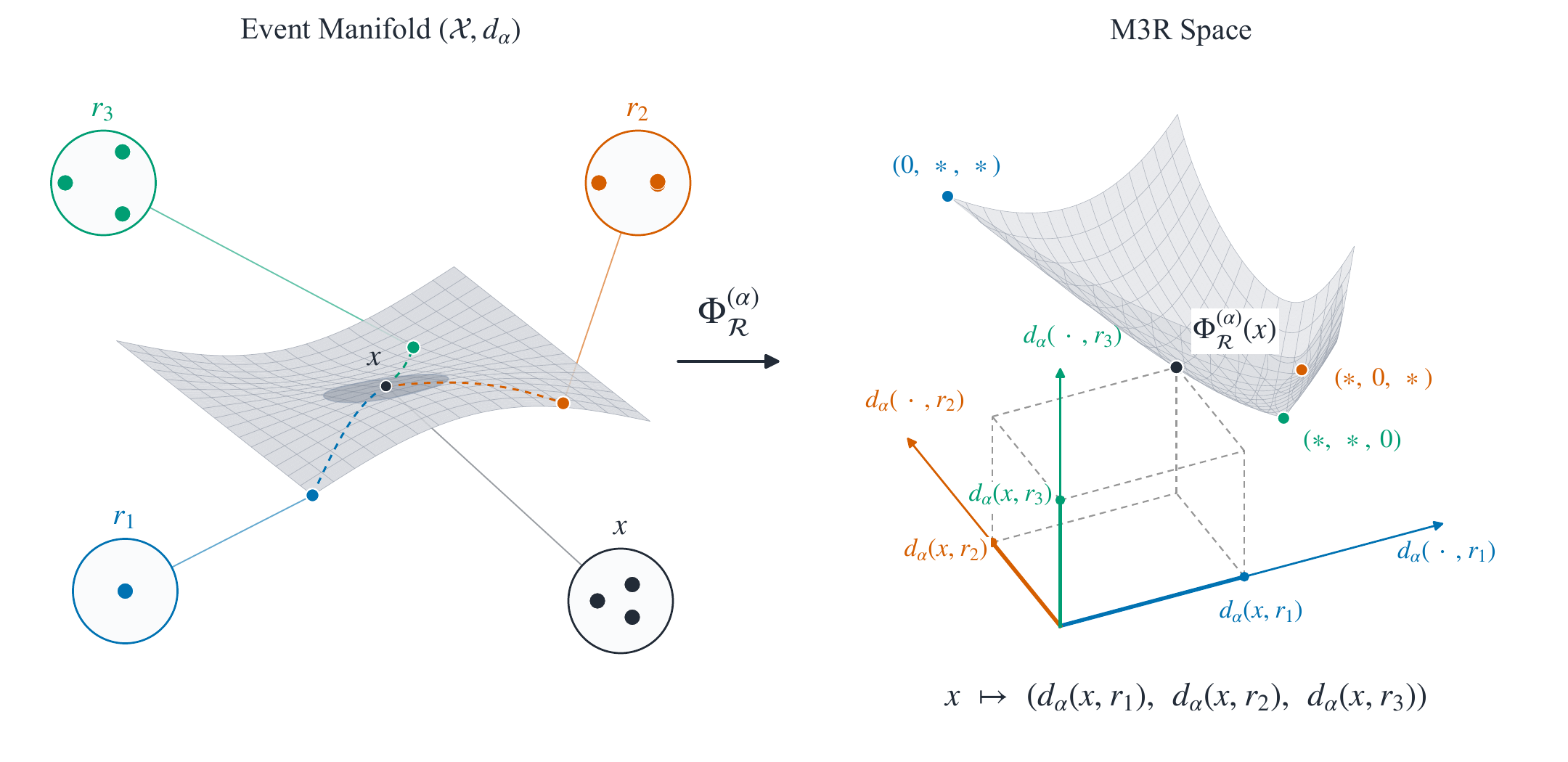}
    \caption{
        Schematic illustration of the M3R construction with three reference events. \textbf{Left:} An event $x$ and three references $(r_1,r_2,r_3)$ in the event manifold $(\mathcal{X},d_\alpha)$. Their pairwise distances to $x$ define its relative coordinates. \textbf{Right:} The corresponding M3R space, in which $x$ is represented by $\Phi_{\mathcal{R}}^{(\alpha)}(x)=(d_\alpha(x,r_1),d_\alpha(x,r_2),d_\alpha(x,r_3))$. The mapped event manifold is shown schematically. The colored points mark the images of the references, $\Phi_{\mathcal{R}}^{(\alpha)}(r_i)$, whose $i$-th coordinate vanishes since $d_\alpha(r_i,r_i)=0$; $\ast$ denotes a generic nonzero value.
    }
    \label{fig:m3r_geom}
\end{figure}

The central idea of M3R is to represent an event relationally with respect to a finite set of reference events. Consider an event space $\mathcal{X}$ equipped with a distance $d_\alpha$, where $\alpha$ labels one of the event metrics introduced in section~\ref{sec:metrics}. We select a reference set
\begin{equation}
    \mathcal{R}
    =
    \{r_1, r_2, \ldots,r_{N_{\rm ref}}\}
    \subset \mathcal{X},
    \label{eq:m3r_reference_set}
\end{equation}
which are randomly sampled from the dataset in this work. An event $x\in\mathcal{X}$ is then mapped to the vector
\begin{equation}
    \Phi_{\mathcal{R}}^{(\alpha)}(x)
    =
    \left(
    d_\alpha(x,r_1),
    d_\alpha(x,r_2),
    \ldots,
    d_\alpha(x,r_{N_{\rm ref}})
    \right)
    \in\mathbb{R}^{N_{\rm ref}}.
    \label{eq:m3r_single_metric}
\end{equation}
Each reference event thus acts as a landmark in the original metric space, while the corresponding distance $d_\alpha(x,r_i)$ provides one coordinate of the relative representation. In other words, rather than requiring an explicit coordinate system for the original metric space, its geometry is now encoded through the relations of each event to a common set of references. Figure~\ref{fig:m3r_geom} illustrates this construction for a toy example with three reference events: each event $x$ is mapped to the point $\Phi_{\mathcal{R}}^{(\alpha)}(x)=(d_\alpha(x,r_1),d_\alpha(x,r_2),d_\alpha(x,r_3))$ in the resulting three-dimensional relative space. Repeating this map over the entire event ensemble produces the corresponding M3R representation of the original event manifold.

For a single underlying metric $d_\alpha$, the Euclidean distance between two relative representations
\begin{align}
    d_{\rm M3R, \mathcal{R}}^{(\alpha)}(x,x’)
    &=
    \left\|
    \Phi_{\mathcal{R}}^{(\alpha)}(x)
    -
    \Phi_{\mathcal{R}}^{(\alpha)}(x’)
    \right\|_2
    \nonumber \\
    &=
    \left(
    \sum_{i=1}^{N_{\rm ref}}
    \left(
    d_\alpha(x,r_i)
    -
    d_\alpha(x’,r_i)
    \right)^2
    \right)^{1/2},
    \label{eq:m3r_distance}
\end{align}
defines the geometry of the corresponding M3R space. The number of reference events $N_{\rm ref}$ controls both the dimensionality of the M3R representation and the amount of relational information retained from the original metric space. A small randomly sampled reference set provides a compressed description based on distances to only a few landmarks, while increasing $N_{\rm ref}$ samples the original geometry more extensively.

The same construction naturally accommodates multiple notions of event similarity. Given a collection of $K$ metrics or dissimilarities $\{d_\alpha\}_{\alpha=1}^{K}$, each defines an individual relative representation $\Phi_{\mathcal{R}}^{(\alpha)}(x)$. These representations can be combined into a single M3R representation
\begin{equation}
    \Phi_{\mathcal{R}}(x)
    =
    \left(
    w_1\Phi_{\mathcal{R}}^{(1)}(x),
    w_2\Phi_{\mathcal{R}}^{(2)}(x),
    \ldots,
    w_K\Phi_{\mathcal{R}}^{(K)}(x)
    \right)
    \in\mathbb{R}^{K \times N_{\rm ref}},
    \label{eq:m3r_multi_metric}
\end{equation}
where the coefficients $w_\alpha$ control the relative normalization of the different metric components. In the numerical analysis, we set $w_1 = \dots = w_K = 1$.  The multi-metric construction therefore places heterogeneous notions of event similarity within a common Euclidean representation without requiring a single metric to be defined directly on the combined space of all event features. In particular, the W$_2$, SW$_2$, and PS geometries can be incorporated alongside the non-kinematic PID-based dissimilarities.

Several useful properties of M3R follow directly from the underlying distance functions. For any reference event $r_i$, the reverse triangle inequality gives
\begin{equation}
    \left|
    d_\alpha(x,r_i)-d_\alpha(x’,r_i)
    \right|
    \leq
    d_\alpha(x,x’),
    \label{eq:m3r_reverse_triangle}
\end{equation}
and therefore from eq.~\ref{eq:m3r_distance} we have
\begin{equation}
    d_{\rm M3R,\mathcal{R}}^{(\alpha)}(x,x’)
    \leq
    \sqrt{N_{\rm ref}}
    d_\alpha(x,x’).
    \label{eq:m3r_lipschitz}
\end{equation}
The relative representation is thus stable under small perturbations in the original metric geometry. A stronger result follows when the entire set of events is used as the reference set. Taking $\mathcal{R}=\mathcal{X}$, eq.~\ref{eq:m3r_reverse_triangle} implies
\begin{equation}
    \left\|
    \Phi_{\mathcal X}^{(\alpha)}(x)
    -\Phi_{\mathcal X}^{(\alpha)}(x')
    \right\|_\infty
    =\sup_{r\in\mathcal{X}}\left|d_\alpha(x,r)-d_\alpha(x',r)\right|
    \leq d_\alpha(x,x').
\end{equation}
Since $x\in\mathcal{X}$ is itself a reference, the coordinate $r=x$ saturates this bound, yielding
\begin{equation}
    \left\|
    \Phi_{\mathcal X}^{(\alpha)}(x)
    -\Phi_{\mathcal X}^{(\alpha)}(x')
    \right\|_\infty
    =d_\alpha(x,x').
    \label{eq:m3r-guarantee-full}
\end{equation}
The complete M3R representation is therefore an isometric embedding of $(\mathcal X,d_\alpha)$ when equipped with the supremum norm and retains the full original metric geometry.

For a finite reference set, however, distinct events need not remain distinguishable. Since $d_{\rm M3R,\mathcal{R}}^{(\alpha)}$ is the Euclidean distance between relative representations, it defines a metric on the image $\Phi_{\mathcal{R}}^{(\alpha)}(\mathcal{X})$. When the reference set is sufficiently discriminating such that
\begin{equation}
    x\neq x’
    \quad\Longrightarrow\quad
    \Phi_{\mathcal{R}}^{(\alpha)}(x)
    \neq
    \Phi_{\mathcal{R}}^{(\alpha)}(x’),
    \label{eq:m3r_injectivity}
\end{equation}
the map is injective and the induced distance also defines a metric on the original event space; otherwise, it defines a pseudometric. Finally, permutations of the reference ordering simply permute the M3R coordinates and leave all Euclidean distances between relative representations unchanged.

For collider applications, the M3R construction also inherits the infrared and collinear behavior of the underlying event distances. If $d_\alpha$ is IRC safe, an infinitesimally soft emission or an exactly collinear splitting of $x$ leaves its distance to every fixed reference event unchanged, i.e.,
\begin{equation}
    d_\alpha(x_{\rm soft/coll.},r_i)
    \rightarrow
    d_\alpha(x,r_i)
    \qquad
    \forall r_i\in\mathcal{R},
    \label{eq:m3r_irc}
\end{equation}
and hence
\begin{equation}
    \Phi_{\mathcal{R}}^{(\alpha)}(x_{\rm soft/coll.})
    \rightarrow
    \Phi_{\mathcal{R}}^{(\alpha)}(x),
  \label{eq:m3r_irc_representation}
\end{equation}
which, again by the reverse triangle inequality, gives
\begin{equation}
    \left|
    d_{\mathcal{R}}(x_{\rm soft/coll.},y)
    -
    d_{\mathcal{R}}(x,y)
    \right|
    \leq
    d_{\mathcal{R}}(x_{\rm soft/coll.},x)
    =
    \left\|
    \Phi_{\mathcal{R}}^{d}(x_{\rm soft/coll.})
    -
    \Phi_{\mathcal{R}}^{d}(x)
    \right\|_2
    \rightarrow 0.
\end{equation}
M3R therefore preserves the IRC behavior of the underlying metrics. Conversely, incorporating IRC-unsafe quantities, such as the PID-based multiplicity distances, retains the corresponding IRC-unsafe property. The framework itself thus imposes no particular notion of physical similarity, but provides a common representation of the information encoded by its constituent distance measures.

Practically, the M3R construction reduces the computational cost of representing large event ensembles. For $N$ events, constructing the complete pairwise distance matrix requires $\mathcal{O}(N^2)$ evaluations of the underlying event metric, whereas constructing the M3R coordinates requires only $\mathcal{O}(N N_{\rm ref})$ event-reference distance evaluations. Once these coordinates are obtained, subsequent comparisons can be performed directly in the finite-dimensional Euclidean representation space. For $N_{\rm ref}\ll N$, M3R therefore provides a computationally efficient compression of the original metric geometry. Operationally, this reduction parallels that achieved by the linearized W$_2$ representation~\cite{PhysRevD.102.116019}, in that $\mathcal{O}(N^2)$ expensive pairwise metric evaluations are replaced by a number of such evaluations linear in $N$ for fixed $N_{\rm ref}$, followed by inexpensive Euclidean comparisons in a finite-dimensional space, although the two constructions are mathematically distinct.

\subsection{Benchmark Dataset and Analysis Setup}
\label{subsec:dataset}

All numerical studies are based on the \texttt{JetClass} dataset~\cite{Qu:2022mxj}, a large-scale public benchmark widely used for developing and evaluating modern machine learning models for jet tagging~\cite{Birk:2023efj, Golling:2024abg, Vigl:2024lat, Brehmer:2024yqw, Mikuni:2025tar, Birk:2024knn, Bardhan:2025icr, Vigl:2026ppx}. \texttt{JetClass} contains ten classes of jets from simulated proton-proton collisions, including light-quark and gluon jets as well as jets arising from top-quark, $W$, $Z$, and Higgs-boson decays, with different decay channels treated as separate classes. This diversity provides a common dataset for studying event geometries across a range of underlying physical processes and jet substructures.

Jets are simulated using \texttt{MadGraph}~\cite{Alwall:2014hca}, \texttt{Pythia}~\cite{Sjostrand:2014zea}, and \texttt{Delphes}~\cite{deFavereau:2013fsa} for hard scattering, parton showering and hadronization, and detector simulation, respectively. They are subsequently reconstructed from detector-level particle-flow objects using the anti-$k_T$ algorithm implemented in \texttt{FastJet}~\cite{Cacciari:2008gp, Cacciari:2011ma} with radius parameter $R=0.8$, and are selected within $500~{\rm GeV}<p_T<1000~{\rm GeV}$ and $|\eta|<2$. The common kinematic selection places the different processes in a similar boosted regime, where the decay products of heavy particles can be mostly contained within a single large-radius jet. Each jet is represented as an unordered set of constituents with variable multiplicity. The available constituent-level features include four-momenta, particle-identification information, electric charge, and trajectory displacement variables for charged particles. In this work, the constituent momenta provide the inputs to the three kinematic metrics introduced in section~\ref{sec:metrics}, while particle identities and electric charges provide the complementary PID information described in section~\ref{subsec:non-kin}.

We select processes spanning several characteristic jet topologies, from predominantly one-pronged QCD jets to two-pronged electroweak and Higgs decays, three-pronged hadronic top decays, and four-pronged Higgs decays. This selection allows us to compare metric geometries across qualitatively different substructures, while the presence of multiple processes with the same nominal prongness enables comparisons between jets with similar coarse topology but different particle content and finer radiation patterns. The specific processes and pairwise classification tasks used in each analysis are introduced in the corresponding sections below.

The entire \texttt{JetClass} dataset contains 100 million training jets, together with additional 5M samples for validation and 20M for testing. Since the analyses in this work require repeated pairwise distance calculations rather than large-scale neural network training, we use $10^4$ jets per selected class throughout. This sample size makes the geometric analyses computationally tractable while allowing the same event ensembles to be compared consistently across different metrics and M3R representations. Our objective is therefore not to optimize classification performance on \texttt{JetClass}, but to characterize the geometric information resolved by different event metrics.

\section{Geometry of Individual Event Manifold} \label{sec:single-manifold}

We begin our geometric analysis by studying individual event manifolds, focusing on how their structure depends on both the metric and the underlying physical process. Section~\ref{subsec:single-manifold-complexity} characterizes their geometric complexity, while section~\ref{subsec:single-manifold-alignment} examines the extent to which manifolds induced by different metrics are geometrically aligned within the common M3R space. These studies of intra-manifold geometry provide the basis for our subsequent analysis of inter-manifold decision boundaries in section~\ref{sec:decision-boundary}.

\subsection{Manifold Complexity}  \label{subsec:single-manifold-complexity}

Several notions of manifold complexity exist, ranging from intrinsic dimensionality to topological and geometric characterizations. Rather than assigning a single mathematical invariant to an event manifold, we adopt an operational notion based on how efficiently the original metric geometry can be reconstructed within M3R. A geometrically simpler manifold should admit a faithful representation using relatively few references, whereas a more intricate geometry may require a larger reference set and retain structures that are difficult to reconstruct. We therefore characterize geometric complexity through the reconstruction fidelity of M3R and its convergence with $N_{\rm ref}$.

Reconstruction fidelity is quantified using two complementary measures. The first probes the preservation of local neighborhoods through a modified Jaccard similarity, defined as the fractional overlap between the $k$-nearest-neighbor sets of each event $x$ in the original metric space and the corresponding M3R space,
\begin{align}
\mathrm{J}_k(x)=
\frac{
\left|\mathrm{KNN}^{d}_k(x)\cap
\mathrm{KNN}^{d_{\rm M3R}}_k(x)\right|
}{
k
} .
\end{align}
Averaging over events and varying $k$ from $1$ to $N-1$, with $N=10^4$, probes neighborhood preservation across geometric scales. We summarize the resulting curve by its normalized area---the Jaccard area under the curve (JAUC)---with values approaching unity for near-complete preservation of the neighborhood structure and $\frac{1}{2}$ for random rankings, analogous to the ROC AUC (see Appendix~\ref{app:methods}).

The second measure, modified average decision ordering (MADO), probes the preservation of relative distance orderings. For pairs of events, it quantifies whether their closer-versus-farther ordering relative to each reference is preserved under the M3R reconstruction, averaged over event pairs and references. MADO approaches unity when these ordering relations are nearly completely preserved. Whereas JAUC measures neighborhood preservation across geometric scales, MADO directly probes pairwise ordering relations and is therefore sensitive to distortions that may leave neighborhood membership unchanged. More details of both measures are provided in Appendix~\ref{app:methods}.

Fig.~\ref{fig:jm_rep} presents reconstruction of the event manifolds as a function of $N_{\rm ref}$ with $1\leq N_{\rm ref}\leq100$. Results are shown for the phase space (PS), spectral 2-Wasserstein (SW$_2$), and energy-flow 2-Wasserstein (W$_2$) metrics across six selected processes in \texttt{JetClass}. The upper panels show the absolute reconstruction fidelity, while the lower panels normalize each curve to its value at $N_{\rm ref}=100$ (denoted as JAUC$_{100}$ or MADO$_{100}$), thereby clearly separating the convergence behavior from differences in the absolute reconstruction level.

\begin{figure}[h]
    \centering
    \includegraphics[width=\linewidth]{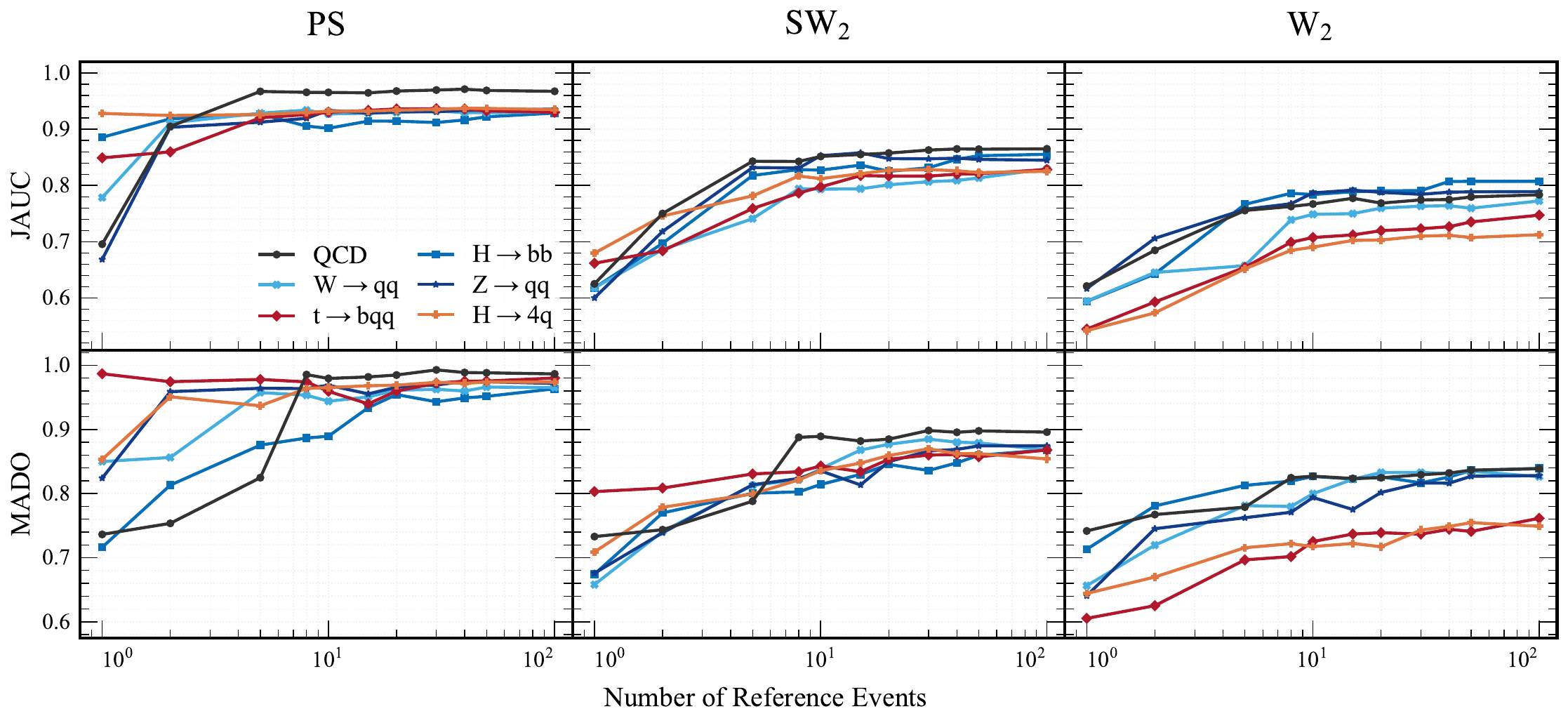}
    \includegraphics[width=\linewidth]{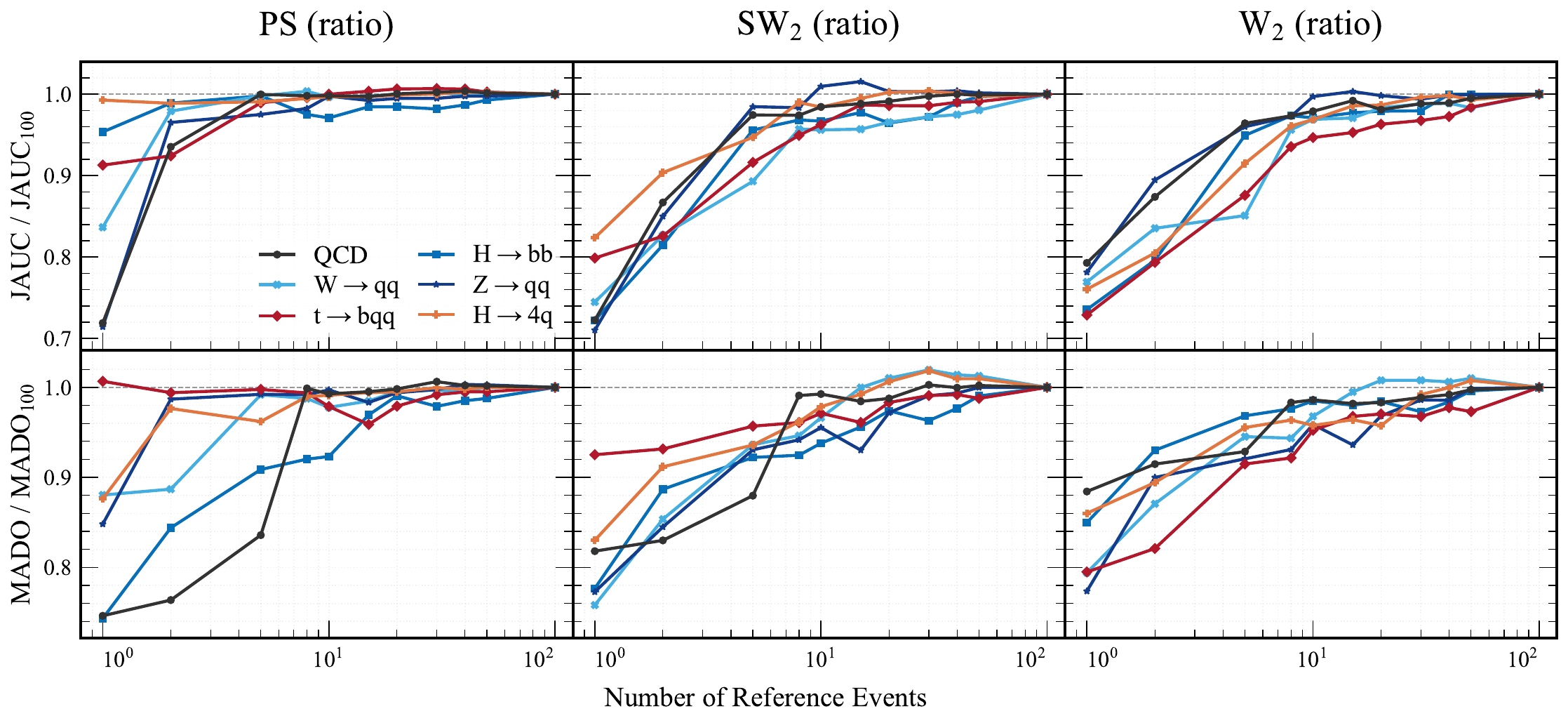}
    \caption{Jaccard area under the curve (JAUC, top) and modified average decision ordering (MADO, bottom) for selected jet classes (colored curves) as functions of $N_{\rm ref}$. Results are shown for three underlying event metrics: the phase space (PS; left), the spectral 2-Wasserstein (SW$_2$; middle), and the 2-Wasserstein (W$_2$; right) distance. The upper panels show the absolute reconstruction fidelity, while the lower panels normalize each curve to its value at $N_{\rm ref}=100$ (denoted as JAUC$_{100}$ or MADO$_{100}$).}
    \label{fig:jm_rep}
\end{figure}

The most prominent feature of Fig.~\ref{fig:jm_rep} is that the manifold reconstruction exhibits a stronger dependence on the event metric than on the underlying physical process. For a fixed metric, the six event classes follow broadly similar reconstruction curves in both JAUC and MADO, particularly at moderate and large $N_{\rm ref}$, indicating comparable geometric complexity as resolved by the two measures. Changing the metric, by contrast, produces qualitatively distinct reconstruction behavior, with sizable differences persisting even as $N_{\rm ref}$ increases. This observation suggests that the geometric complexity of an event manifold is governed primarily by the notion of similarity used to resolve the event space, rather than by the particular physical process populating it. In this sense, PS, SW$_2$, and W$_2$ do not merely provide alternative numerical measures of separation between the same events; they induce geometrically distinct organizations of a given event ensemble, with each exhibiting substantial commonality across the processes considered.

Process-dependent effects nevertheless remain as a secondary source of variation, with the spread among event classes being most pronounced for W$_2$ ($\sim10\%$). This is qualitatively consistent with the different resolving powers of the metrics: metrics sensitive to finer variations in radiation patterns can expose process-specific structures that are less apparent under coarser descriptions. For W$_2$, for example, the QCD manifold is comparatively easier to reconstruct within the M3R framework, while reconstruction becomes more challenging for structured multi-pronged processes such as $t\to bqq$ and $H\to4q$. Process-dependent variations are also visible for SW$_2$ and PS, although the ordering among event classes is less pronounced and less systematic. The residual process dependence should therefore be understood as an interplay between the complexity of the underlying event structure and the extent to which a given metric resolves it, rather than as an intrinsic hierarchy among event classes.

More specifically, the $N_{\rm ref}$ dependence reveals two distinct aspects of manifold complexity. We define the \textit{reconstruction complexity} through the absolute reconstruction fidelity attained at large $N_{\rm ref}$\footnote{We verified the stability of the large-$N_{\rm ref}$ reconstruction level by extending the reference set up to $N_{\rm ref}=200$. The reconstruction measures exhibit no further increase, confirming that $N_{\rm ref}=100$ is already representative of the saturated regime for the manifolds considered here.}, which characterizes how much of the original metric geometry can ultimately be captured within the M3R space. By contrast, the \textit{convergence complexity} characterizes how rapidly this attainable reconstruction is recovered as $N_{\rm ref}$ increases, and therefore the compressibility of the manifold under a finite-reference representation. In other words, reconstruction complexity quantifies how much of the original geometry can be reconstructed, while convergence complexity quantifies how efficiently that reconstruction can be achieved with a finite reference set. The latter is more directly exposed by the normalized curves in the lower panels of Fig.~\ref{fig:jm_rep}, where differences in the absolute reconstruction level have been factored out.

Turning first to the convergence complexity, the normalized curves demonstrate that all event manifolds considered here are highly compressible under the M3R representation. The largest improvement occurs at small $N_{\rm ref}$, followed by a rapid approach to the large-$N_{\rm ref}$ behavior. For PS, most of the reconstruction achieved at $N_{\rm ref}=100$ is already recovered with $\mathcal{O}(10)$ reference events. The convergence is more gradual for SW$_2$ and W$_2$, but likewise approaches their respective $N_{\rm ref}=100$ levels with only a few tens of references. This is orders of magnitude smaller than the $10^4$ events sampling each process, indicating that most of the geometric structure accessible to the M3R representation can be recovered from only a small fraction of the event samples. Moreover, for a fixed metric, the normalized curves remain broadly similar across event classes, indicating that the convergence complexity is only weakly process dependent.

A markedly stronger metric dependence emerges in the reconstruction complexity, reflected by the absolute reconstruction level reached at large $N_{\rm ref}$. At $N_{\rm ref}=100$, PS achieves JAUC values above $0.92$ and MADO values above $0.96$ across all processes considered, corresponding to near-complete preservation of the local neighborhood structure and relative ordering of pairwise distances. For QCD in particular, JAUC and MADO reach up to $0.97$ and $0.99$, respectively. The corresponding JAUC (MADO) values range from approximately $0.83$--$0.87$ ($0.85$--$0.90$) for $\mathrm{SW}_2$ and $0.71$--$0.81$ ($0.75$--$0.84$) for $\mathrm{W}_2$, establishing a clear hierarchy in reconstruction complexity. Here again, W$_2$ exhibits the largest process-to-process variation, although this variation remains subleading to the differences between metrics.

The hierarchy is consistent with the different levels of physical information resolved by the three metrics. The PS construction reduces each event to a fixed number of reconstructed objects ($N=10$ here) and then compares their $N$-body phase-space configurations, thereby integrating out much of the soft and fine-grained constituent-level radiation. By contrast, SW$_2$ retains constituent-level information through the event spectral function, while W$_2$ acts directly on the energy flow and is sensitive to the most detailed variations in the radiation pattern. The resulting hierarchy in reconstruction complexity,
\begin{align}
\text{PS} < \text{SW}_2 < \text{W}_2,
\end{align}
is therefore consistent with the increasing granularity at which these metrics resolve collider events. Within M3R, the finer structure retained by W$_2$ is correspondingly more difficult to reconstruct from a finite set of references, whereas the more compressed PS geometry can be recovered substantially more completely. The larger process-to-process variation observed for W$_2$ is also consistent with the same picture, as its finer resolution exposes more process-specific radiation structure.

The analysis above establishes that different metrics induce event manifolds with different reconstruction complexities, reflecting the level of geometric structure resolved by each metric. It alone, however, does not determine whether these manifolds encode distinct geometries or share a common structure supplemented by metric-specific information. The common M3R space allows this question to be addressed directly by examining the extent to which manifolds induced by different metrics align for the same physical process.

\subsection{Manifold Alignment}  \label{subsec:single-manifold-alignment}

To investigate the alignment between event manifolds, a principal component analysis (PCA) is performed on the M3R spaces induced by PS, SW$_2$, and W$_2$ for three representative processes: QCD, $W\to qq$, and $t\to bqq$. These processes are selected to span qualitatively different jet topologies, from one-pronged QCD radiation to characteristic two- and three-pronged structures. For each physical process, the same ensemble of $10^4$ jets is represented under the three metrics using a common set of $N_{\rm ref}=20$ reference events. The resulting M3R spaces therefore share the same coordinate basis, allowing their principal directions to be compared directly.

\begin{figure}[h]
    \centering
    \includegraphics[width=1\linewidth]{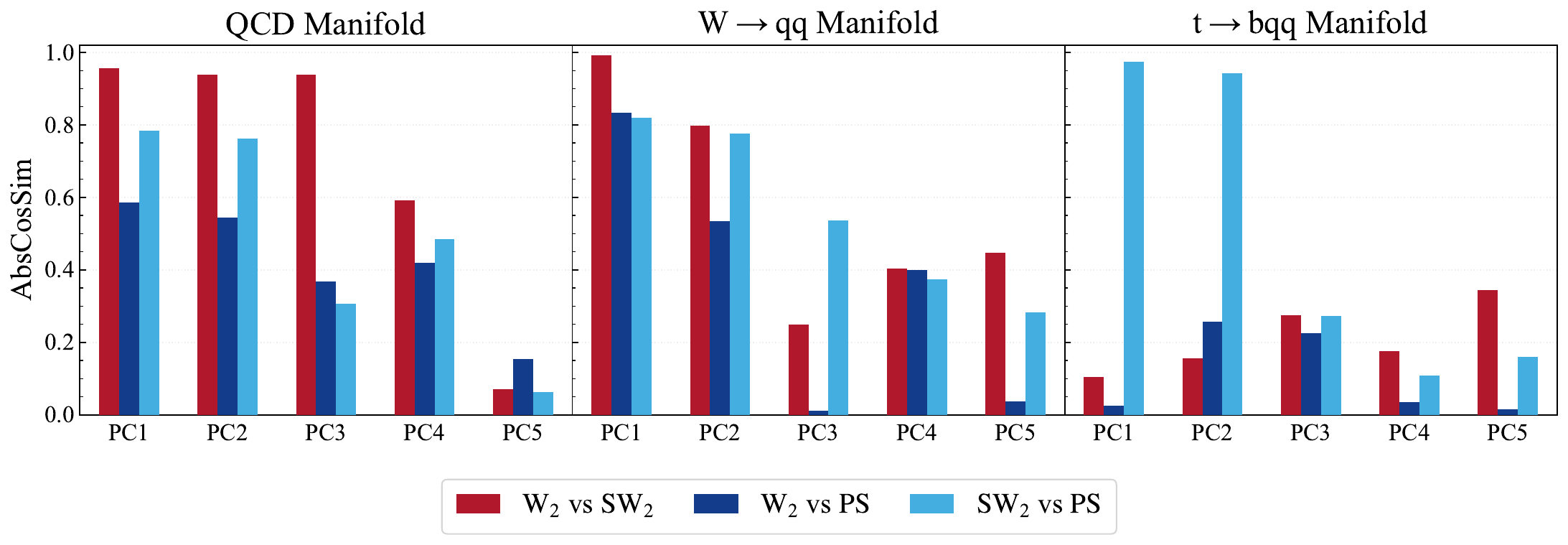}
    \caption{Absolute cosine similarity between the first five principal component directions of the M3R representations (with $N_{\rm ref}=20$) of 10k QCD jets (left), 10k $W\to qq$ jets (middle), and 10k $t\to bqq$ jets (right). The three bars compare the PCA directions of the event manifolds induced by different metrics: W$_2$ vs.~SW$_2$ (red), W$_2$ vs.~PS (dark blue), and SW$_2$ vs.~PS (light blue). Larger values indicate stronger alignment between the principal components.}
    \label{fig:qcd_pca_alignment}
\end{figure}

For two metrics $d_\alpha$ and $d_\beta$, the alignment of their $i$-th principal directions is quantified by the absolute cosine similarity
\begin{align}
\operatorname{AbsCosSim}_i^{(\alpha,\beta)}
=\left|
\hat{\mathbf v}^{(\alpha)}_i\cdot
\hat{\mathbf v}^{(\beta)}_i
\right|,
\end{align}
where $\hat{\mathbf v}^{(\alpha)}_i$ denotes the normalized $i$-th principal direction of the manifold induced by metric $d_\alpha$. The absolute value accounts for the arbitrary sign of PCA eigenvectors, such that $\operatorname{AbsCosSim}_i^{(\alpha,\beta)}=1$ corresponds to coincident principal axes, while $0$ indicates orthogonal directions. Figure~\ref{fig:qcd_pca_alignment} shows the resulting alignment for the first five principal directions of each pair of metric-induced manifolds.

A general feature of Fig.~\ref{fig:qcd_pca_alignment} is that the strongest cross-metric alignment tends to occur among the leading principal directions. For several metric pairs, the first one or two components exhibit $\operatorname{AbsCosSim}$ values close to unity, while the alignment generally becomes much weaker and less systematic for higher principal components. This trend is, however, not universal, and the detailed manifold alignment pattern varies substantially across physical processes, both in the relative alignment among metric pairs and in its evolution across successive principal components.

For QCD jets, the three metric-induced manifolds exhibit substantial alignment in their leading principal directions, with the strongest correspondence observed between W$_2$ and SW$_2$. In particular, their first three principal axes reach $\operatorname{AbsCosSim}\simeq0.95$, while appreciable alignment ($\gtrsim 0.5$) persists into the subleading components. This behavior is consistent with the known perturbative correspondence between the energy-flow and spectral Wasserstein metrics. In the soft and collinear limits relevant for QCD radiation, the two constructions share the same leading behavior, and their two-point event correlators have recently been shown to coincide through next-to-next-to-leading logarithmic accuracy up to the appropriate normalization~\cite{6r43-xhtj}. Our results provide a further manifestation of this correspondence in the geometry of the full simulated event manifold, beyond the perturbative observables for which the analytical relation has been established. The common soft-collinear structure dominating QCD radiation is reflected in the near alignment of the leading principal directions independently induced by SW$_2$ and W$_2$. The alignment is nevertheless not exact and decreases along the subleading directions, consistent with the two metrics inducing distinct geometries away from the perturbative limits in which their equivalence is established. More broadly, PS also exhibits strong alignment with SW$_2$ and moderate alignment with W$_2$ along the first two principal directions, suggesting that the dominant QCD geometry is partially shared across all three metric constructions.

A different alignment pattern emerges for the $W\to qq$ manifold. The first principal direction is nearly coincident across all three metrics, with $\operatorname{AbsCosSim}\gtrsim0.8$ for every metric pair. This remarkably strong three-way alignment indicates that the dominant variation of the $W$ event manifold is resolved similarly by PS, SW$_2$, and W$_2$, reflecting a common geometric mode associated with the characteristic two-prong structure. The correspondence, however, deteriorates rapidly beyond the leading directions. At PC3, SW$_2$ and PS retain moderate alignment ($\operatorname{AbsCosSim}\simeq0.55$), whereas the alignment of W$_2$ with either metric falls below $0.25$ (close to 0 for W$_2$ and PS). The $W\to qq$ manifold therefore provides a particularly clear example in which a strongly shared dominant geometric direction coexists with substantially metric-dependent subleading structure.

The $t\to bqq$ manifold exhibits the strongest metric dependence among the three processes considered. In particular, W$_2$ becomes weakly aligned with both SW$_2$ and PS already in the first two principal directions ($\operatorname{AbsCosSim}\lesssim0.25$), while SW$_2$ and PS are extremely well-aligned at $\simeq0.95$. This behavior differs qualitatively from both QCD and $W\to qq$, and is consistent with the theoretical analysis~\cite{Gambhir:2024ndc}. For QCD jets, emissions from the hard core are typically strongly ordered in energy and angle, suppressing configurations for which distinct energy-flow geometries become degenerate in the spectral representation. Top jets, by contrast, naturally contain three hard sub-prongs forming nontrivial triangular configurations. Their spatial arrangement is resolved directly by W$_2$, whereas the spectral representation can map geometrically distinct configurations onto similar spectral functions, making the differences between the two metrics more pronounced. The present results provide a manifold-level manifestation of this distinction, with the dominant W$_2$ and SW$_2$ directions becoming only weakly aligned in the common M3R space.

Across all three processes, SW$_2$ and PS exhibit substantial alignment in their leading principal directions, suggesting a comparatively process-robust component shared by the two metric-induced geometries. This similarity is nontrivial given their rather different constructions. One possibility is that their leading principal directions are governed by a common set of coarse kinematic variations that both metrics resolve efficiently, while their differences reside predominantly in finer geometric structures. Establishing the physical origin of this persistent alignment, for example by relating the principal directions to explicit kinematic observables, is left for future investigation.

No universal hierarchy of cross-metric alignment persists across physical processes, in contrast to the reconstruction complexity in section~\ref{subsec:single-manifold-complexity}, for which metric dependence dominates over process dependence and leads to the approximately universal ordering $\text{PS} < \text{SW}_2 < \text{W}_2$. Manifold alignment therefore exhibits a substantially stronger dependence on the interplay between the metric and the underlying physical process than does the complexity of each manifold considered individually.

\begin{figure}[h]
    \centering
    \includegraphics[width=1\linewidth]{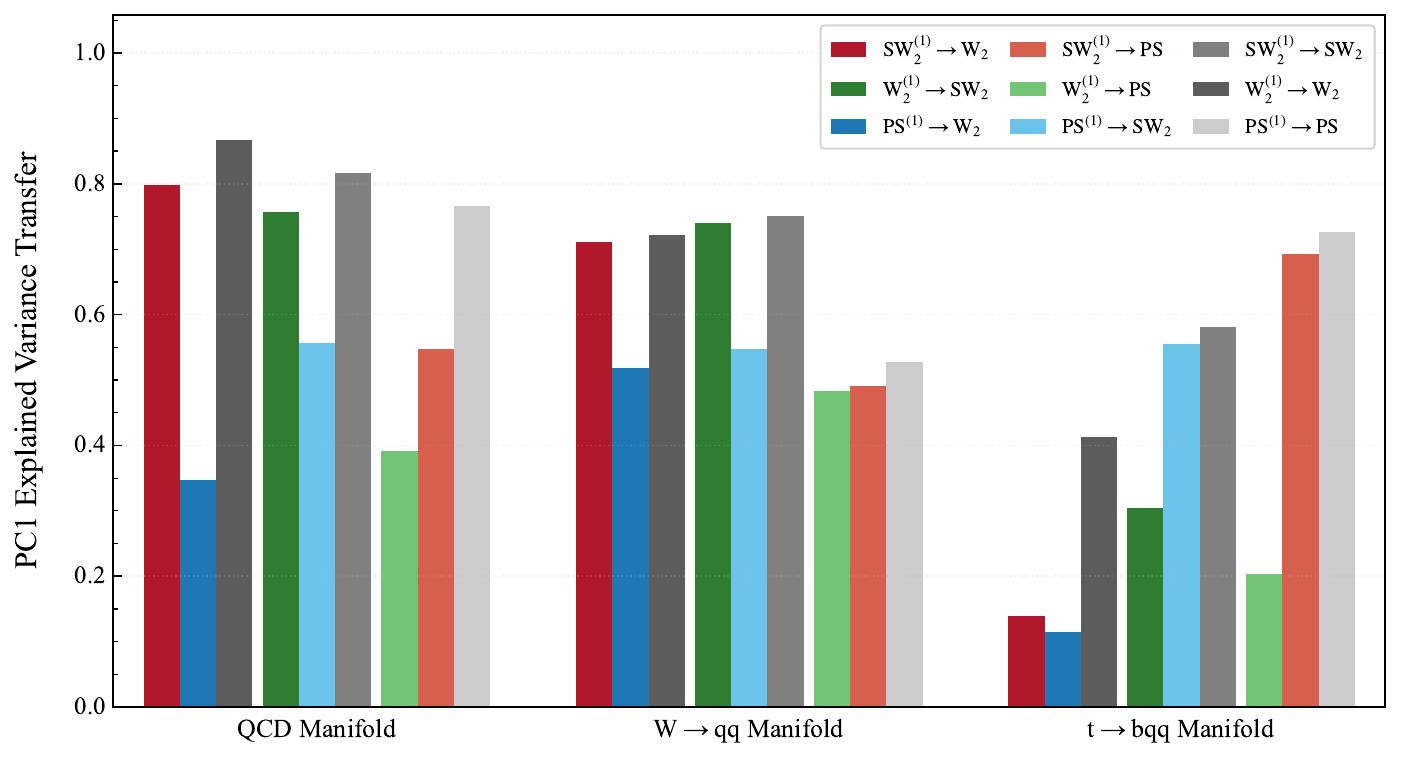}
    \caption{Fraction of variance in the target metric-induced M3R manifold explained by projecting onto the first principal component (PC1) direction obtained from a source metric-induced manifold, for QCD, $W\to qq$, and $t\to bqq$ jets. For each jet type, the six colored bars correspond to all directed transfers among W$_2$, SW$_2$, and PS, grouped by target metric and colored by source metric. The gray bars show the self-transfers, which give the maximum target-manifold variance attainable by a single linear direction.}
    \label{fig:ev_trans_pc1}
\end{figure}

The analysis above quantifies the relative orientation of principal directions, but does not determine whether a direction important to one manifold captures a significant fraction of the variation in another. To address this question, Fig.~\ref{fig:ev_trans_pc1} considers the transfer of the leading principal direction between metric-induced M3R manifolds. For a source metric $d_\alpha$ and target metric $d_\beta$, the fraction of the target-manifold variance explained by the first principal direction of the source manifold is defined as
\begin{equation}
\operatorname{EV}\left(\alpha^{(1)}\to \beta\right)
=
\frac{
\operatorname{Var}\left[\mathbf{X}^{(\beta)}\hat{\mathbf v}^{(\alpha)}_1\right]
}{
\sum_i \lambda^{(\beta)}_i
},
\label{eq:ev_transfer}
\end{equation}
where $\mathbf{X}^{(\beta)}$ is the mean-centered M3R data matrix of the jets under the target metric $d_\beta$, $\hat{\mathbf v}^{(\alpha)}_1$ is the first principal direction of the source manifold, and the denominator is the total variance of the target manifold. Fig.~\ref{fig:ev_trans_pc1} shows all directed transfers among PS, SW$_2$, and W$_2$; for example, $\text{SW}_2^{(1)}\to \text{W}_2$ denotes the fraction of W$_2$-manifold variance explained by projection onto the first principal direction of the SW$_2$ manifold. The corresponding self-projection, $\beta^{(1)}\to \beta$, gives the variance explained by the target manifold's own PC1 and therefore provides the natural reference for the maximum variance attainable by a single linear direction.

The variance-transfer results corroborate and sharpen the alignment patterns observed in Fig.~\ref{fig:qcd_pca_alignment}. For QCD, the leading W$_2$ and SW$_2$ directions show substantial mutual transfer: $\text{SW}_2^{(1)}$ explains approximately $0.8$ of the variance in the W$_2$ manifold, compared with $\simeq0.87$ for the W$_2$ self-PC1, which by construction represents the maximum variance attainable by any one-dimensional linear projection. A similarly strong transfer is observed in the reverse direction. The correspondence is even more striking for $W\to qq$, where transfers between the two Wasserstein manifolds and from either Wasserstein manifold to PS approach their respective self-PC1 values. The reverse transfers from PS are notably weaker. This indicates that, in most cases, a principal direction extracted from one manifold captures nearly as much of the target-manifold variation as its own optimal direction. The nearly coincident PC1 directions identified in Fig.~\ref{fig:qcd_pca_alignment} therefore reflect not only similar orientations, but a genuinely shared dominant mode of manifold variation. By contrast, for the $t\to bqq$ manifold, SW$_2^{(1)}$ and PS$^{(1)}$ each explain only $\lesssim0.15$ of the W$_2$ variance, compared with $\sim0.4$ for W$_2^{(1)}$, while substantially more variance is transferred between SW$_2$ and PS. The weak alignment with W$_2$ observed in Fig.~\ref{fig:qcd_pca_alignment} thus reflects not merely a difference in principal-axis orientation, but a genuinely distinct dominant mode of variation in the W$_2$ manifold.

The cross-metric variance transfer is furthermore directional: in general, $\operatorname{EV}(\alpha^{(1)}\to \beta)\neq\operatorname{EV}(\beta^{(1)}\to \alpha)$. For example, for the $t\to bqq$ manifold, the W$_2$ PC1 explains approximately $0.3$ of the variance in SW$_2$, while the reverse transfers account for less than $0.15$ of the W$_2$ variance. The asymmetry persists after normalization
to the respective self-PC1 values. Combined with the greater reconstruction complexity of W$_2$ found in section~\ref{subsec:single-manifold-complexity}, this behavior is consistent with a picture in which W$_2$ resolves both a common geometric structure and additional metric-specific information. Establishing such a directional hierarchy more generally, however, requires extending the analysis beyond the leading principal direction and is left for future work.

Sections~\ref{subsec:single-manifold-complexity} and~\ref{subsec:single-manifold-alignment} provide a complementary characterization of the intra-manifold geometry associated with an individual physical process. The complexity of each event manifold is shown to be governed predominantly by the metric, while the geometric relations among different metric-induced manifolds exhibit a stronger dependence on the underlying physical process, with both shared and metric-specific structures emerging in the common M3R space. Having established these properties within individual event manifolds, the next question is how manifolds associated with distinct physical processes are geometrically separated. Section~\ref{sec:decision-boundary} therefore turns to the inter-manifold geometry and investigates the decision boundaries between different event classes.

\section{Geometry of Inter-Manifold Decision Boundary} \label{sec:decision-boundary}

We study the geometry separating different event manifolds using support vector machines (SVMs)~\cite{Cortes:1995hrp} as classifiers. For two classes, an SVM constructs a maximum-margin decision boundary.~\footnote{In the soft-margin formulation used here, violations of the margin are allowed with a penalty controlled by the hyperparameter $C$. Larger $C$ penalizes margin violations more strongly, while smaller $C$ allows a wider margin.} With the radial basis function (RBF) kernel used here, the decision function can be written as
\begin{equation}
    f(x)
    =
    \sum_{n\in{\rm SV}}
    a_n y_n
    \exp\left[
        -\gamma
        \left(
        d_{\rm M3R,\mathcal{R}}^{(\alpha)}(x,x_n)
        \right)^2
    \right]
    +b,
    \label{eq:svm_decision_function}
\end{equation}
where $y_n=\pm1$ labels the class of support vector $x_n$, the weights $a_n$ and offset $b$ are determined during training, and the sign of $f(x)$ determines the predicted class of event $x$.~\footnote{Support vectors are training events that lie on or within the margin, or on the wrong side of it. The RBF hyperparameter $\gamma$ sets the characteristic distance scale of the kernel, with larger $\gamma$ allowing more irregular boundaries. All hyperparameters are fixed throughout the analyses in this section.}

The SVM construction provides two complementary probes of the inter-manifold geometry particularly useful for our purposes. First, because the kernel depends on events only through their Euclidean M3R distance, the classification performance quantifies how faithfully M3R preserves the decision-relevant information of the original metric geometry. Second, for a fixed SVM prescription, the fraction of training events that become support vectors provides an operational measure of decision-boundary complexity, with a larger support vector fraction indicating a more complex boundary. Section~\ref{subsec:decision-boundary-preserve} first studies the preservation of decision-relevant geometry under M3R, while section~\ref{subsec:decision-boundary-complexity} examines the complexity of the resulting decision boundaries.

\subsection{Preservation of Decision-relevant Geometry} \label{subsec:decision-boundary-preserve}

\begin{figure}[h]
    \centering
    \includegraphics[width=\linewidth]{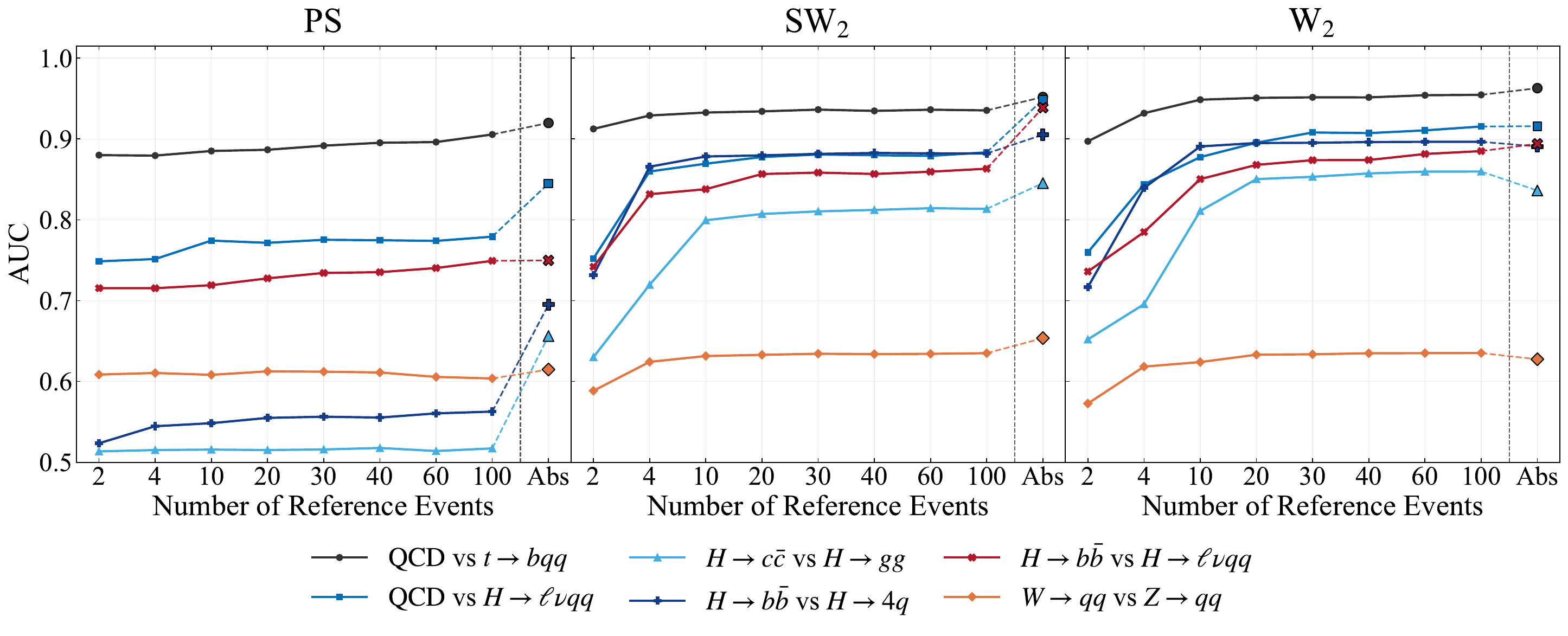}
    \caption{SVM classification performance as a function of the number of reference events $N_{\rm ref}$ for the PS (left), SW$_2$ (center), and W$_2$ (right) relative representations, for six selected binary classification tasks. The points labeled ``Abs'' show the corresponding AUCs obtained directly using the original metrics.}
    \label{fig:svm_curves_all}
\end{figure}

Figure~\ref{fig:svm_curves_all} examines how effectively the M3R representations induced by different metrics preserve the geometric information relevant for separating physical processes. We consider six tagging tasks spanning a broad range of separability, from the well distinguishable QCD \textit{vs.}~$t\to bqq$ classification to the more challenging discrimination between $W\to qq$ and $Z\to qq$. For each binary tagging task, an SVM classifier with an RBF kernel is trained on the M3R representations constructed from PS, SW$_2$, and W$_2$, with the number of reference events varied from $N_{\rm ref}=2$ to $100$ drawn equally from the two jet classes. As a benchmark, we train the same SVM directly in the original metric space (denoted as ``Abs'').~\footnote{In this case, the kernel in eq.~\ref{eq:svm_decision_function} is evaluated with $d_\alpha(x,x_n)$ in place of $d_{\rm M3R,\mathcal{R}}^{(\alpha)}(x,x_n)$.}

In general, M3R succeeds in reproducing the tagging performance of the original metric with a substantially smaller number of reference events. For most metric--task combinations, the AUC obtained at $N_{\rm ref}\sim10$ already approaches the corresponding ``Abs'' result, with the notable exceptions of several PS-based tasks to be discussed below. This immediately provides a computational advantage: constructing the full metric representation requires $\mathcal{O}(N_{\rm evt}^2)$ pairwise distance evaluations and storage, whereas M3R requires only $\mathcal{O}(N_{\rm evt}N_{\rm ref})$. For $N_{\rm ref}\ll N_{\rm evt}$, M3R can therefore preserve approximately the same tagging performance while substantially reducing the computational cost. This extends, in a metric-agnostic and multi-reference setting, the practical motivation underlying the linearized W$_2$ construction~\cite{PhysRevD.102.116019}, where a single-reference representation similarly enabled substantial computational saving while retaining comparable discrimination performance.

Fig.~\ref{fig:svm_curves_all} reveals contrasts with the intra-manifold reconstruction studied in section~\ref{subsec:single-manifold-complexity}. Despite exhibiting the greatest reconstruction complexity, W$_2$ shows the most complete preservation of tagging performance: at $N_{\rm ref}=100$, its AUC closely matches the corresponding ``Abs'' value across all six tasks, with an average relative difference of approximately $-0.5\%$, compared with $4\%$ for SW$_2$ and $9\%$ for PS. The resulting pattern is therefore the reverse of the reconstruction complexity hierarchy ($\mathrm{PS}<\mathrm{SW}_2<\mathrm{W}_2$) found for individual event manifolds, showing that single-manifold reconstruction and preservation of classification performance probe different aspects of the underlying metric geometry. Faithful reconstruction of the full metric geometry does not seem to be necessary to preserve the structures relevant for discrimination. Section~\ref{subsec:decision-boundary-complexity} examines whether this reversed hierarchy is also reflected in the complexity of the corresponding decision boundaries.

The preservation of decision-relevant geometry depends strongly on the particular processes being separated. QCD \textit{vs.}~$t$, for example, is an intrinsically simple tagging task and its AUCs reach $\sim 0.9$ under all three metrics even with only two references. A largely consistent hierarchy of tagging difficulty emerges across the metrics: QCD \textit{vs.}~$t\to bqq$ is followed by QCD \textit{vs.}~$H\to \ell\nu qq$, while the Higgs--Higgs tasks are generally more challenging and $W\to qq$ \textit{vs.}~$Z\to qq$ is among the most difficult. This ordering broadly reflects the expected physical similarity of the corresponding processes, and is particularly consistent for W$_2$ and SW$_2$. PS presents notable exceptions, particularly for $H\to c\bar c$ \textit{vs.}~$H\to gg$ and $H\to b\bar b$ \textit{vs.}~$H\to 4q$, where the finite-reference representation retains a sizable gap from the corresponding ``Abs'' performance even at large $N_{\rm ref}$.

We have therefore established that the geometry required for discrimination can be different from the full geometry of the individual event manifolds, with their respective metric orderings being seemingly reversed. Classification performance alone, however, does not reveal the geometric complexity of the boundary separating two manifolds. This raises a further question: does the reversed metric ordering observed here also reflect a corresponding hierarchy in the complexity of the decision boundaries themselves?

\subsection{Complexity of Decision Boundary} \label{subsec:decision-boundary-complexity}

\begin{figure}[h]
    \centering
    \includegraphics[width=1\linewidth]{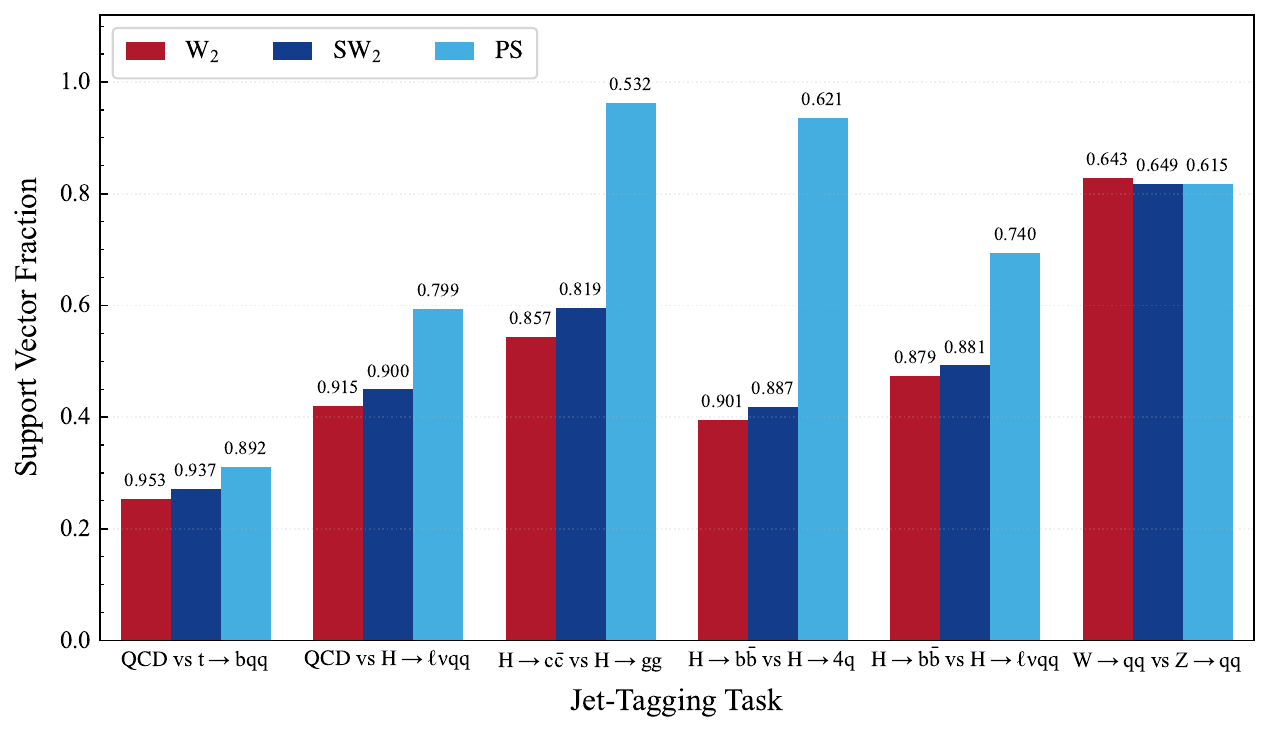}
    \caption{Support vector fractions for six binary jet-tagging tasks using SVM classifiers trained on the relative representations with $N_{\rm ref}=20$ (10 from each class). Results are shown for the three metrics: W$_2$, SW$_2$, and PS. For all metrics and tagging tasks, the SVM hyperparameters are fixed to $\gamma=\texttt{scale}$ \protect\footnotemark{} and $C=10$. The support vector fraction measures the fraction of training events retained as support vectors, and the number at the top of each bar shows the corresponding tagging AUC.}
    \label{fig:rel_sv_fractions}
\end{figure}

To probe the complexity of the inter-manifold decision boundary, we use the fraction of SVM training events retained as support vectors as an operational geometric diagnostic. Analogous to the manifold complexity introduced in section~\ref{subsec:single-manifold-complexity}, this notion of \textit{decision boundary complexity} is defined through the resources required to represent the corresponding geometry, rather than through an intrinsic geometric invariant. Importantly, M3R places the different metric-induced geometries in a common $N_{\rm ref}$-dimensional space, allowing their support vector fractions to be compared on equal footing. Fig.~\ref{fig:rel_sv_fractions} shows this quantity for the six tagging tasks using $N_{\rm ref}=20$.~\footnote{The choice of $N_{\rm ref}$ is motivated by section~\ref{subsec:decision-boundary-preserve}, where $\mathcal{O}(10)$ references were found sufficient to preserve most of the decision-relevant information.}~\footnotetext{\texttt{scale} is the default $\gamma$ value for the RBF kernel in \texttt{scikit-learn}~\cite{Pedregosa:2011ork}. The precise definition can be found at \href{https://scikit-learn.org/stable/modules/generated/sklearn.svm.SVC.html\#sklearn.svm.SVC}{\texttt{sklearn.svm.SVC}}.}

We again observe the strong process dependence seen in Fig.~\ref{fig:svm_curves_all}, now manifested directly in the complexity of the inter-manifold decision boundary. For QCD \textit{vs.}~$t$, less than one-third of the training events are retained as support vectors across all three metrics, whereas $W\to qq$ \textit{vs.}~$Z\to qq$ requires approximately 80\% for all metrics, with the remaining tasks generally interpolating between these limits. This hierarchy closely follows the tagging difficulty identified in the previous subsection, shown here by the corresponding AUC at $N_{\rm ref}=20$.  The two quantities, however, characterize related but distinct properties: AUC measures the separability of the two manifolds, while the support vector fraction quantifies the sparsity of the kernel expansion defining the learned SVM decision function. Fig.~\ref{fig:rel_sv_fractions} thus extends the performance-based characterization of Fig.~\ref{fig:svm_curves_all} to the geometry of the separating boundary itself. Processes that are more readily separable generally also admit sparser, and operationally simpler, inter-manifold decision boundaries.

At fixed process pair, Fig.~\ref{fig:rel_sv_fractions} reveals a notable metric dependence in the decision boundary complexity. The differences are most pronounced for the Higgs–Higgs tagging tasks, where PS requires a substantially larger fraction of support vectors than the two Wasserstein metrics. For $H\to c\bar c$ \textit{vs.}~$H\to gg$, for example, less than 60\% of the training events are retained as support vectors for W$_2$ and SW$_2$, compared with nearly all events for PS. More generally, W$_2$ and SW$_2$ tend to induce inter-manifold decision boundaries of comparable complexity, both typically simpler than those induced by PS.

This reversed tendency highlights a fundamental distinction between the individual event manifold (section~\ref{sec:single-manifold}) and the inter-manifold decision boundary (section~\ref{sec:decision-boundary}): the two are different geometric objects and need not exhibit the same complexity. A finer-grained metric such as W$_2$ resolves more constituent-level radiation structure within each process, producing a richer individual manifold that is more difficult for M3R to reconstruct. However, when this additional structure corresponds to physical degrees of freedom along which the processes differ, it can also produce a cleaner geometric separation and a decision boundary requiring fewer support vectors. This effect may be particularly important for closely related processes such as Higgs--Higgs tagging, where discrimination relies on subtle differences in radiation and substructure. While W$_2$ and SW$_2$ may resolve these differences as separation between the manifolds, the reduced PS description may partially collapse the corresponding discriminating directions, leaving the two manifolds more intertwined. A metric can therefore induce a more complex individual manifold while simultaneously producing a simpler inter-manifold decision boundary.

\section{From Multiple Metrics to a Unified Representation} \label{sec:unified-manifolds}

Sections~\ref{sec:single-manifold} and~\ref{sec:decision-boundary} have shown that different collider metrics resolve distinct but partially overlapping structures, both within individual event manifolds and in the boundaries separating physical processes. This raises a natural question: can these complementary notions of event similarity be combined to improve event discrimination beyond what is achievable with any individual metric? M3R provides a natural framework for this purpose, allowing relative representations derived from different metrics, together with non-kinematic information, to be combined at the event level.

We construct the unified representation by concatenating the M3R coordinates introduced in section~\ref{sec:m3r},
\begin{align}
    \Phi_{\mathcal R}^{\rm uni}(x)= \Phi_{\mathcal R}^{\rm PS}(x) \oplus \Phi_{\mathcal R}^{\rm SW_2}(x) \oplus \Phi_{\mathcal R}^{\rm W_2}(x) \oplus \Phi_{\mathcal R}^{\rm PID}(x),
\label{eq:unified_m3r}
\end{align}
where $\oplus$ denotes feature concatenation and $\Phi_{\mathcal R}^{\rm PID}$ incorporates the non-kinematic information defined in eq.~\ref{eq:nonkinematic_distance}. We use the same $N_{\rm ref}=40$ reference events (20 from each class) for all $\Phi_{\mathcal R}$'s. By successively adding these feature blocks, we can directly quantify the incremental information contributed by each notion of event similarity.

\begin{figure}[h]
    \centering
    \includegraphics[width=1\linewidth]{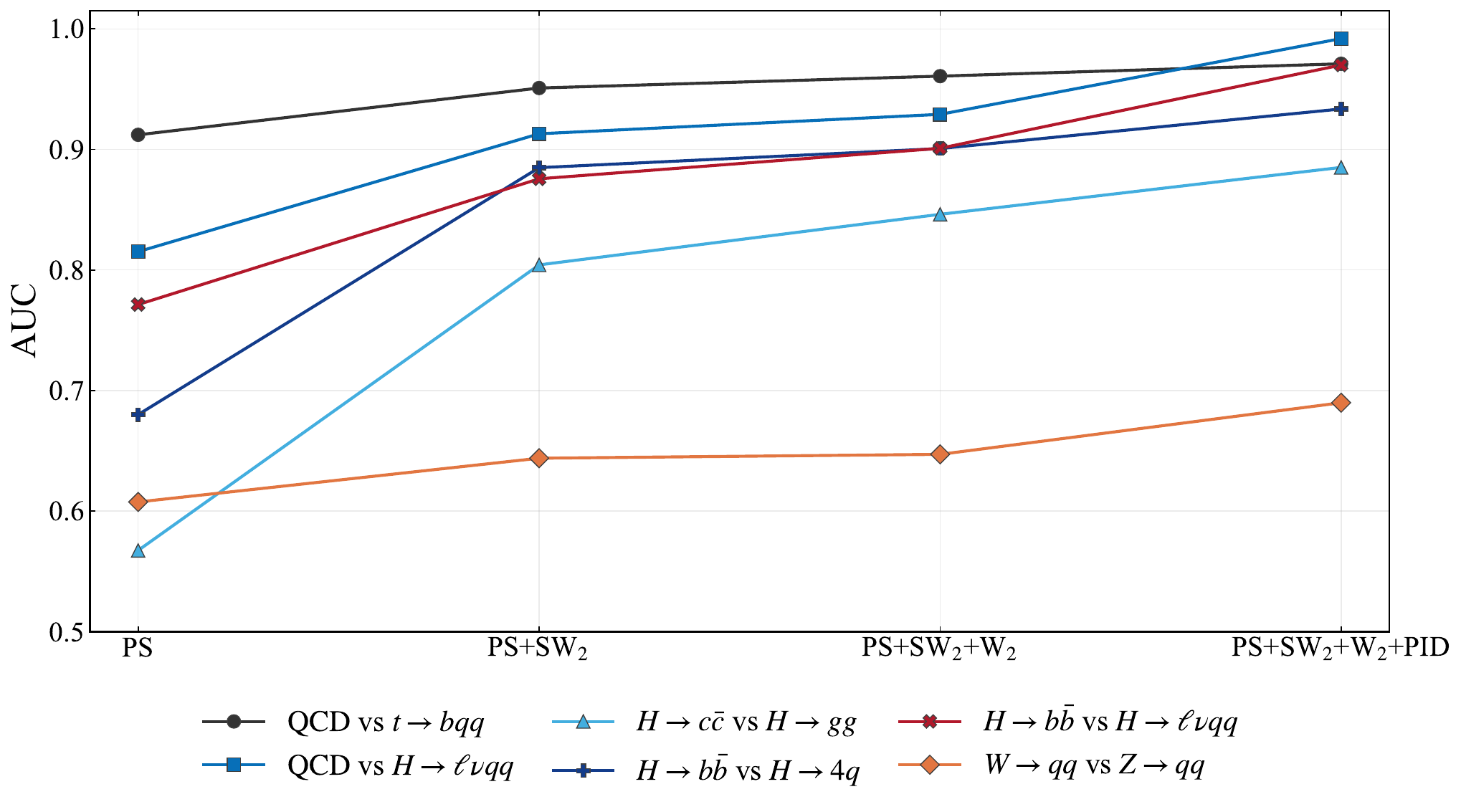}
    \caption{
    BDT classification AUCs obtained by successively combining the PS, SW$_2$, W$_2$, and non-kinematic PID relative representations within the M3R framework, for selected pairs of physical processes. Each point includes all feature blocks to its left, illustrating the incremental information contributed by different notions of event similarity.}
    \label{fig:unified_repr}
\end{figure}

Because the M3R feature blocks differ in dimensionality and numerical scale, with each kinematic metric contributing $N_{\rm ref}$ coordinates and PID contributing $6N_{\rm ref}$, we use a boosted decision tree (BDT)~\cite{Friedman:2001wbq} to combine them for classification.~\footnote{Specifically, we use the histogram-based gradient-boosted decision tree (\texttt{HistGradient BoostingClassifier}) implemented in \texttt{scikit-learn}~\cite{Pedregosa:2011ork}, trained with binary cross-entropy loss and early stopping. We use a learning rate of $0.05$, at most $31$ leaves per tree, at least $50$ events per leaf, an $L_2$ regularization strength of $1$, and $255$ histogram bins per feature. Training is allowed to proceed for up to $2000$ boosting iterations and is terminated if the validation loss does not improve for $20$ consecutive iterations.} This avoids imposing a predefined weighting between feature blocks and allows their complementary information to be incorporated directly by the classifier.~\footnote{As a robustness check, we additionally tested a multi-tower multilayer perceptron (MLP), in which the individual M3R feature blocks are first processed by separate network branches and then combined through a joint representation with learned relative weighting. The same qualitative trends were observed, while the BDT generally achieved better classification performance.} For each binary tagging task, the BDT is trained with separate training, validation, and test samples, with the validation sample used for early stopping.~\footnote{For each task, the $2\times10^4$ events are stratified by class and divided into training, validation, and test samples in the proportions $72\%$, $8\%$, and $20\%$, respectively.} Performance is quantified by the AUC on the independent test sample. Figure~\ref{fig:unified_repr} and Table~\ref{tab:feature_progression} show the classification performance as the individual M3R feature blocks are successively combined. Figure~\ref{fig:unified_repr} follows the progression PS $\rightarrow$ SW$_2$ $\rightarrow$ W$_2$ $\rightarrow$ PID, while Table~\ref{tab:feature_progression} additionally considers the ordering SW$_2$ $\rightarrow$ W$_2$ $\rightarrow$ PS $\rightarrow$ PID. Comparing the two orderings helps disentangle the incremental contribution of each feature block and reveals the complementarity and redundancy among different notions of event similarity.

\begin{table}[h]
  \centering
  \small
  \renewcommand{\arraystretch}{1.15}
  \setlength{\tabcolsep}{4pt}
  \begin{tabular}{@{}|l|c|c|c|c|||c|c|c|c|@{}}
    \hline
    & \(\mathrm{PS}\) & \(+\mathrm{SW}_2\) & \(+\mathrm{W}_2\) & \(+\mathrm{PID}\) & \(\mathrm{SW}_2\) & \(+\mathrm{W}_2\) & \(+\mathrm{PS}\) & \(+\mathrm{PID}\) \\
    \hline
    \(\mathrm{QCD}\) vs\ \(t\to bqq\) & 0.9121 & 0.9509 & 0.9608 & 0.9710 & 0.9493 & 0.9611 & 0.9608 & 0.9710 \\
    \(\mathrm{QCD}\) vs\ \(H \to \ell\nu q q\) & 0.8152 & \textbf{\underline{0.9130}} & 0.9291 & \textbf{0.9919} & 0.8929 & 0.9255 & 0.9291 & \textbf{0.9919} \\
    \(H \to c\bar{c}\) vs\ \(H \to gg\) & 0.5674 & \textbf{\underline{0.8042}} & \textbf{0.8462} & 0.8851 & 0.7994 & \textbf{0.8460} & 0.8462 & 0.8851 \\
    \(H \to b\bar{b}\) vs\ \(H \to 4q\) & 0.6800 & \textbf{\underline{0.8849}} & 0.9007 & 0.9335 & 0.8841 & 0.9021 & 0.9007 & 0.9335 \\
    \(H \to b\bar{b}\) vs\ \(H \to \ell \nu q q\) & 0.7713 & \textbf{\underline{0.8756}} & 0.9009 & \textbf{0.9700} & 0.8585 & 0.8953 & 0.9009 & \textbf{0.9700} \\
    \(W \to qq\) vs\ \(Z \to qq\) & 0.6076 & \textbf{0.6439} & 0.6471 & \textbf{0.6899} & 0.6440 & 0.6500 & 0.6471 & \textbf{0.6899} \\
    \hline
  \end{tabular}
  \caption{BDT classification AUCs for two orderings in which the M3R feature blocks are successively combined. Within each ordering, a column labeled \(+X\) reports the AUC after adding feature block \(X\) to those preceding it. Bold entries indicate an incremental AUC improvement greater than 5\%, while bold and underlined entries indicate an improvement greater than 10\%.}
  \label{tab:feature_progression}
\end{table}

The successive addition of M3R feature blocks generally improves classification performance. For the ordering in Fig.~\ref{fig:unified_repr}, adding SW$_2$ to PS increases the AUC across all six tagging tasks, with further gains from W$_2$ and PID. Such improvement is not guaranteed simply by introducing additional features: heterogeneous representations may contain redundant or competing information that provides little additional discriminative power and can even degrade classification performance. The observed gains therefore indicate that the distinct geometries identified in sections~\ref{sec:single-manifold} and~\ref{sec:decision-boundary} contain complementary information that can be jointly exploited within the unified M3R representation.

The gains from combining multiple relative representations are strongly task dependent. For the simplest QCD \textit{vs.}~$t$ tagging task, individual metrics already achieve AUC $>0.9$, leaving little room for improvement (though improvements still exist). The gains are substantially larger for the more challenging Higgs--Higgs discrimination tasks, where no single metric fully captures the relevant discriminative information. For $H\to c\bar c$ \textit{vs.}~$H\to gg$, for example, the AUC increases from approximately $0.57$ with PS alone to $0.8$ after adding SW$_2$, and to nearly $0.9$ with W$_2$ and PID. The contribution from W$_2$ remains substantial under the alternative ordering in Table~\ref{tab:feature_progression}, where adding W$_2$ to SW$_2$ increases the AUC by more than 5\%. The complementarity among metric-induced geometries is therefore most consequential when separating similar processes requires sensitivity to multiple aspects of the underlying event structure.

Adding PID information provides the largest improvements ($>5\%$) for $W\to qq$ \textit{vs.}~$Z\to qq$, $H\to b\bar b$ \textit{vs.}~$H\to \ell \nu qq$, and QCD \textit{vs.}~$H\to \ell \nu qq$. While hadronic $W$ and $Z$ decays exhibit similar two-prong kinematics, they differ in their electric charge and quark-flavor composition. The semileptonic Higgs decay, in turn, is distinguished from predominantly hadronic processes by the presence of a prompt charged lepton. These physical differences are reflected in the final-state particle composition and charge, allowing PID to provide complementary information beyond kinematics. M3R can therefore effectively combine not only distinct kinematic geometries but also qualitatively different, non-kinematic notions of event similarity.

The two feature-block orderings in Table~\ref{tab:feature_progression} further reveal an asymmetric redundancy among the metric-induced representations. Starting from PS, adding SW$_2$ and W$_2$ can substantially improve the AUC, whereas adding PS after SW$_2$ and W$_2$ yields little further improvement across the tagging tasks. This suggests that much of the discriminative information captured by the reduced PS geometry is already accessible to the constituent-level Wasserstein representations, while SW$_2$ and W$_2$ encode additional information not generally captured by PS.

\section{Conclusions and Outlook} \label{sec:conclusions}

We introduced the Multi-Reference Relative Representation (M3R) as a unified framework for studying collider event geometries through distances to a finite set of reference events. Focusing on the phase space (PS), spectral 2-Wasserstein (SW$_2$), and energy-flow 2-Wasserstein (W$_2$) metrics, we investigated three related questions: how metrics and physical processes shape individual event manifolds; how intra-manifold structure relates to inter-manifold geometry separating different processes; and how complementary metric geometries can be combined into a unified event representation. Our results reveal several general features of the geometric organization of collider event space.

First, individual event manifold geometry is governed more strongly by the defining metric than by the underlying physical process: reconstruction behavior is broadly similar across processes for a fixed metric, while substantially larger differences arise between metrics. Our operational characterization further distinguishes two aspects of manifold complexity. Convergence complexity is generally small, with most accessible geometric information recovered using only $\mathcal{O}(10)$ references, whereas reconstruction complexity exhibits a robust hierarchy, $\mathrm{PS}<\mathrm{SW}_2<\mathrm{W}_2$, consistent with the progressively finer event structure resolved by these metrics.

Second, different metrics resolve both shared and metric-specific structures within the same event manifold, with their correspondence depending on the underlying physical process. Their leading M3R principal directions can exhibit substantial cross-metric alignment, while subleading structures are generally more distinct. For QCD jets, W$_2$ and SW$_2$ align strongly along their dominant directions, consistent with their known correspondence in the soft-collinear regime; for multi-pronged top jets, this alignment breaks down and W$_2$ develops a distinct dominant structure. Cross-metric alignment is therefore not a property of the metrics alone, but also of the physical events on which they act.

Third, the geometry relevant for distinguishing physical processes can differ substantially from the geometry within individual event manifolds. Most decision-relevant information is captured with relatively few references, even when the underlying manifold geometry remains only partially reconstructed. Moreover, the metric hierarchy is approximately reversed: while W$_2$ induces the richest and most difficult-to-reconstruct individual manifolds, its decision-relevant information is generally best preserved by M3R, and the Wasserstein metrics tend to yield simpler SVM decision boundaries than PS for most tagging tasks. Intra- and inter-manifold complexity are therefore distinct geometric properties: a metric can resolve richer structure within individual processes while simultaneously providing a simpler separation between them.

Finally, the distinct metric geometries contain complementary information that can be exploited jointly. Combining their M3R representations generally preserves or improves discrimination, with the largest gains for more challenging tasks. This complementarity is asymmetric: SW$_2$ and W$_2$ provide substantial additional information beyond PS, whereas PS contributes little in reverse, suggesting that much of its decision-relevant information is already captured by the Wasserstein metrics. The complementarity remains process dependent and extends beyond kinematics, as demonstrated by the additional gains from PID information for processes such as $W/Z$ discrimination.

These results open several directions for further developing the geometric description. An important step is to connect the operational notions of geometric complexity developed here with intrinsic mathematical and information-theoretic characterizations. While our M3R-based definitions directly probe event manifold reconstruction and discrimination, complementary quantities such as intrinsic dimension, metric entropy, and geometric measures of manifold separation may capture other aspects of the same event spaces. Information geometry provides a related perspective on statistical distinguishability through quantities such as Fisher information and statistical divergences~\cite{kuntz2026latentinformationgeometryjet}. Relating these concepts to the reconstruction and discrimination complexities studied here may help explain the observed hierarchies and connect the empirical geometry of collider events to its underlying mathematical and statistical structure.

A more fundamental goal is to develop a systematic and physically grounded understanding of collider event geometry. The observed alignment between W$_2$ and SW$_2$ for QCD jets is consistent with their known correspondence in the soft-collinear regime, while their divergence for top jets reflects their different sensitivity to more complex event structures. Establishing such connections systematically may explain when metric geometries align or diverge, the origin of their complexity hierarchies, and how their complementary information should be combined. Conversely, a more principled understanding of event manifolds may turn geometry itself into a probe of collision dynamics, revealing structures associated with hard scattering, perturbative QCD radiation, and non-perturbative effects such as hadronization.

Beyond these conceptual developments, several methodological extensions could improve the framework. While reference events are randomly sampled in this work, adaptive or data-driven selection strategies could identify more informative reference sets, improving reconstruction fidelity at smaller $N_{\rm ref}$. The selected references may themselves carry physical information by identifying regions of the event manifold most important for reconstructing its geometry. Another extension is to incorporate richer notions of event similarity. Rather than combining non-kinematic information such as PID only at the representation level, it could be incorporated directly into the underlying event metric, for example through multispecies optimal transport.

The analysis can also be extended beyond the Standard Model jet types in \texttt{JetClass}, particularly to BSM signatures with qualitatively different radiation patterns and event topologies. One example is the high-multiplicity, quasi-isotropic Hidden Valley scenario~\cite{STRASSLER2007374, Han_2008}, for which the phase space metric has shown strong discrimination from QCD dijet events~\cite{Cai2025phasespace}. Such processes provide a natural test of whether the geometric hierarchies observed here persist across qualitatively different collider signatures. Further, it would be interesting to study how these geometries evolve under increasingly realistic event descriptions, including variations in parton showers and hadronization modeling, detector effects, and pileup. This would help disentangle structures associated with the underlying hard scattering from those introduced or modified by subsequent modeling and experimental effects, and ultimately assess their robustness in collider data.

Another promising direction is to extend this geometric perspective to representations learned by modern collider machine learning models. These models implicitly organize events in their latent spaces, yet the relation between these learned geometries and physically defined event metrics remains largely unexplored. A common geometric language could reveal whether learned representations recover structures associated with known metrics, combine multiple notions of similarity, or develop qualitatively distinct geometries. Physically motivated metrics could thereby serve as probes of how machine learning models organize event information. Exploring these connections within and beyond M3R is the subject of ongoing work.

More broadly, the M3R framework suggests moving beyond a single prescribed notion of collider event geometry toward a picture in which multiple geometries coexist and contribute according to the problem at hand. From this perspective, collider event geometry is no longer a fixed structure imposed on data, but a compositional object shaped jointly by the underlying physics and the notions of similarity through which it is resolved.

The code used in this work is publicly available at \url{https://github.com/TianjiCai/collider-metric-m3r}.

\acknowledgments

We would like to thank David Shih, Sung Hak Lim, and Won Sang Cho for useful discussions. The work of TC is supported by the Fundamental Research Funds for the Central Universities and sponsored by Shanghai Pujiang Programme. The authors acknowledge the use of OpenAI models for assistance with manuscript editing and code development, and Anthropic Claude for assistance with code development. All scientific content, analyses, and conclusions were developed and verified by the authors.

\appendix

\section{Additional Details}\label{app:methods}

\subsection*{Modified Jaccard AUC}
\label{app:jauc proof}

We derive the limiting behavior of the modified Jaccard AUC (JAUC) used in section~\ref{sec:single-manifold}. For a query event $x$ drawn from a sample of $N$ events, let $A_k(x)$ and $B_k(x)$ denote its sets of $k$
nearest neighbors under two different distance measures $d_\alpha$ and $d_\beta$, respectively, with $x$ itself excluded. The modified Jaccard similarity is defined as
\begin{equation}
    J_k(x)
    =
    \frac{|A_k(x)\cap B_k(x)|}{k},
\end{equation}
which measures the fraction of neighbors shared between the two $k$-nearest-neighbor sets. There are $M=N-1$ possible neighbors for each query event; any distance ties are resolved using a fixed ordering.

Consider first the null hypothesis in which the two distance measures induce independent, uniformly random neighbor rankings. A neighborhood containing $k$ events occupies a fraction $k/M$ of all possible
neighbors. Therefore, if the two rankings are unrelated, each member of $A_k(x)$ has probability $k/M$ of also appearing in $B_k(x)$. The expected number of shared neighbors is consequently $k^2/M$, giving
\begin{equation}
    \mathbb{E}[J_k(x)]
    =
    \frac{k^2/M}{k}
    =
    \frac{k}{M}
    =
    \frac{k}{N-1}
    \simeq
    \frac{k}{N}.
    \label{eq:knn-overlap-null}
\end{equation}
Introducing the normalized neighborhood size
\begin{equation}
    t=\frac{k}{M},
\end{equation}
the expected overlap for random rankings therefore follows the diagonal, $\mathbb{E}[J_k]=t$. At the opposite limit, if the two distance measures induce identical neighbor rankings, then $A_k(x)=B_k(x)$ for every $k$,
and hence $J_k(x)=1$.

Averaging over all query events, we define
\begin{equation}
    \bar J_k
    =
    \frac{1}{N}\sum_x J_k(x).
\end{equation}
Since every query event has the same expected random overlap $k/M$, the event average obeys
\begin{equation}
    \mathbb{E}[\bar J_k]
    =
    \frac{k}{M}
    =
    t.
\end{equation}
Thus, averaging over events does not change the limiting behavior: random rankings follow the diagonal $\bar J=t$, while identical rankings give $\bar J=1$.

Finally, we summarize the neighborhood agreement across all scales by the discrete modified Jaccard area under the curve (JAUC),
\begin{equation}
    \mathrm{JAUC}_M
    =
    \frac{1}{M}
    \sum_{k=1}^{M}\bar J_k.
\end{equation}
For random rankings, this is the discrete area under the diagonal,
\begin{equation}
    \mathbb{E}[\mathrm{JAUC}_M]_{\rm random}
    =
    \frac{1}{M}
    \sum_{k=1}^{M}\frac{k}{M}
    =
    \frac{M+1}{2M}
    \xrightarrow[M\to\infty]{}
    \frac{1}{2},
    \label{eq:knn-overlap-area-random}
\end{equation}
whereas identical rankings give
\begin{equation}
    \mathrm{JAUC}_{\rm identical}=1.
    \label{eq:knn-overlap-area-identical}
\end{equation}
Geometrically, these limits correspond to the area under the diagonal $\bar J=t$ and the horizontal line $\bar J=1$, respectively. The JAUC therefore has an asymptotic random-ranking baseline of $1/2$ and reaches
unity when the neighborhood structure is perfectly preserved.

\subsection*{Modified Average Decision Ordering}
\label{app:mado}

The modified average decision ordering (MADO) provides a complementary measure of how faithfully the M3R representation preserves the relative distance orderings of the original metric geometry. Consider two events
$x$ and $x'$ and a probe event $s_j$ drawn from the event sample $\mathcal{S}=\{s_1,\ldots,s_N\}$. The original metric $d_\alpha$ determines which of $x$ and $x'$ lies closer to $s_j$, and the M3R distance $d_{\rm M3R,\mathcal{R}}^{(\alpha)}$ makes the same comparison in the relative representation. The two geometries agree whenever
\begin{equation}
    \left[d_\alpha(x,s_j)-d_\alpha(x',s_j)\right]
    \left[
    d_{\rm M3R,\mathcal{R}}^{(\alpha)}(x,s_j)
    -d_{\rm M3R,\mathcal{R}}^{(\alpha)}(x',s_j)
    \right]
    >0,
\end{equation}
that is, when both identify the same event as closer to the probe.

For a given pair $(x,x')$, we define the decision-ordering agreement as the fraction of probe events for which this ordering is preserved,
\begin{equation}
    \mathrm{DO}^{(\alpha)}_{\mathcal R}(x,x')
    =
    \frac{1}{N}\sum_{j=1}^{N}
    \Theta\!\left(
    \left[d_\alpha(x,s_j)-d_\alpha(x',s_j)\right]
    \left[
    d_{\rm M3R,\mathcal{R}}^{(\alpha)}(x,s_j)
    -d_{\rm M3R,\mathcal{R}}^{(\alpha)}(x',s_j)
    \right]
    \right),
    \label{eq:do-pair}
\end{equation}
where $\Theta$ is the Heaviside step function with $\Theta(0)=1/2$, assigning the neutral value $\frac{1}{2}$ whenever either distance gives a tie. Averaging this quantity over all unordered pairs of events gives
\begin{equation}
    \mathrm{MADO}^{(\alpha)}_{\mathcal R}
    =
    \frac{2}{N(N-1)}
    \sum_{a<b}
    \mathrm{DO}^{(\alpha)}_{\mathcal R}(s_a,s_b).
    \label{eq:mado}
\end{equation}

The limiting behavior has a simple interpretation. If the M3R reconstruction perfectly preserves the relative distance ordering of the original metric, every probe makes the same closer-versus-farther decision and $\mathrm{MADO}=1$. If instead the two orderings are statistically unrelated, either event is equally likely to be identified as closer to a generic probe, giving
\begin{equation}
    \mathbb{E}[\mathrm{MADO}]_{\rm random}
    \simeq \frac{1}{2},
\end{equation}
up to negligible finite-sample corrections. MADO therefore has a random-ordering baseline of approximately $\frac{1}{2}$, while values approaching unity indicate increasingly faithful preservation of the relative distance ordering of the original event geometry.

In practice, evaluating MADO over all event pairs and probe events would be computationally expensive for large event samples. We therefore estimate the averages by Monte Carlo sampling, using $2\times10^4$ randomly selected event pairs and $128$ randomly selected probe events. The same pairs and probes are used for all values of $N_{\rm ref}$ to ensure a consistent comparison, while the probes are sampled independently of the reference set $\mathcal{R}$ used to construct the M3R representation.


\bibliographystyle{JHEP}
\bibliography{draft}

\end{document}